\documentclass[acmsmall,xcolor={table}]{acmart}
\pdfoutput=1 %

\usepackage{listings}

\definecolor{mygreen}{rgb}{0,0.6,0}
\definecolor{mygray}{rgb}{0.5,0.5,0.5}
\definecolor{mymauve}{rgb}{0.58,0,0.82}

\lstdefinestyle{PythonStyle}{ 
  backgroundcolor=\color{white},   %
  basicstyle=\scriptsize,        %
  breakatwhitespace=false,         %
  breaklines=true,                 %
  captionpos=b,                    %
  commentstyle=\color{mygreen},    %
  deletekeywords={...},            %
  escapeinside={\%*}{*)},          %
  extendedchars=true,              %
  frame=single,	                   %
  keepspaces=true,                 %
  keywordstyle=\color{blue},       %
  language=Python,                 %
  morekeywords={*,...},            %
  numbers=left,                    %
  numbersep=5pt,                   %
  numberstyle=\tiny\color{mygray}, %
  rulecolor=\color{black},         %
  showspaces=false,                %
  showstringspaces=false,          %
  showtabs=false,                  %
  stepnumber=1,                    %
  stringstyle=\color{mymauve},     %
  tabsize=2,	                   %
  title=\lstname                   %
}

\lstdefinestyle{TextStyle}{ 
  backgroundcolor=\color{white},   %
  basicstyle=\scriptsize,        %
  breakatwhitespace=false,         %
  breaklines=true,                 %
  captionpos=b,                    %
  extendedchars=true,              %
  frame=single,	                   %
  keepspaces=true,                 %
  rulecolor=\color{black},         %
  showspaces=false,                %
  showstringspaces=false,          %
  showtabs=false,                  %
  stepnumber=1,                    %
  tabsize=2,	                   %
  title=\lstname                   %
}

\definecolor{codegreen}{rgb}{0,0.6,0}
\definecolor{codegray}{rgb}{0.5,0.5,0.5}
\definecolor{codepurple}{rgb}{0.58,0,0.82}
\definecolor{backcolour}{rgb}{0.95,0.95,0.92}

\lstdefinestyle{mystyle}{
    backgroundcolor=\color{backcolour},   
    commentstyle=\color{codegreen},
    keywordstyle=\color{magenta},
    numberstyle=\tiny\color{codegray},
    stringstyle=\color{codepurple},
    basicstyle=\ttfamily\footnotesize,
    breakatwhitespace=false,         
    breaklines=true,                 
    captionpos=b,                    
    keepspaces=true,                 
    numbers=left,                    
    numbersep=5pt,                  
    showspaces=false,                
    showstringspaces=false,
    showtabs=false,                  
    tabsize=2
}

\AtBeginDocument{%
  \providecommand\BibTeX{{%
    \normalfont B\kern-0.5em{\scshape i\kern-0.25em b}\kern-0.8em\TeX}}}

\setcopyright{acmlicensed}
\acmJournal{TECS}
\acmYear{2023} \acmVolume{1} \acmNumber{1} \acmArticle{1} \acmMonth{1} \acmPrice{15.00}\acmDOI{10.1145/3577200}

\usepackage{pifont}
\usepackage{framed}
\usepackage{soul}
\usepackage{booktabs}  
\usepackage{csquotes}
\usepackage{multirow}
\usepackage{array}
\usepackage{subcaption}

\usepackage[linesnumbered,vlined,boxed,ruled]{algorithm2e}

\SetKwInput{KwInput}{Input}                %
\SetKwInput{KwOutput}{Output}              %

\newcolumntype{L}[1]{>{\raggedright\let\newline\\\arraybackslash\hspace{0pt}}m{#1}}
\newcolumntype{C}[1]{>{\centering\let\newline\\\arraybackslash\hspace{0pt}}m{#1}}
\newcolumntype{R}[1]{>{\raggedleft\let\newline\\\arraybackslash\hspace{0pt}}m{#1}}
\usepackage{url}
\MakeOuterQuote{"}

\newcommand{\emphbox}[1]{\noindent\definecolor{shadecolor}{RGB}{244, 233, 247}\fcolorbox{black}{shadecolor}{\parbox{0.99\textwidth}{#1}}}

\newcommand{\clrpg}{}

\begin{document}

\title{High-Level Approaches to Hardware Security: A Tutorial}

\author{Hammond Pearce}
\authornote{The authors contributed equally to this work.}
\email{hammond.pearce@nyu.edu}
\orcid{0000-0002-3488-7004}
\affiliation{%
  \institution{New York University}
  \streetaddress{6 MetroTech Center}
  \city{Brooklyn}
  \state{New York}
  \country{USA}
  \postcode{11201}
}

\author{Ramesh Karri}
\email{rkarri@nyu.edu}
\orcid{0000-0001-7989-5617}
\authornotemark[1]
\affiliation{%
  \institution{New York University}
  \streetaddress{6 MetroTech Center}
  \city{Brooklyn}
  \state{New York}
  \country{USA}
  \postcode{11201}
}

\author{Benjamin Tan}
\email{benjamin.tan1@ucalgary.ca}
\orcid{0000-0002-7642-3638}
\authornotemark[1]
\affiliation{%
  \institution{University of Calgary}
  \streetaddress{2500 University Drive NW}
  \city{Calgary}
  \state{Alberta}
  \country{Canada}
  \postcode{T2N 1N4}
}

\renewcommand{\shortauthors}{Pearce, Karri, and Tan}

\begin{abstract}
Designers use third-party intellectual property (IP) cores and outsource various steps in the integrated circuit (IC) design and manufacturing flow. 
As a result, security vulnerabilities have been rising. 
This is forcing IC designers and end users to re-evaluate their trust in ICs. If attackers get hold of an unprotected IC, they can reverse engineer the IC and pirate the IP.  
Similarly, if attackers get hold of a design, they can insert malicious circuits or take advantage of ``backdoors'' in a design. 
Unintended design bugs can also result in security weaknesses.
This tutorial paper provides an introduction to the domain of hardware security through two pedagogical examples of hardware security problems. 
The first is a walk-through of the scan chain-based side channel attack. 
The second is a walk-through of logic locking of digital designs. 
The tutorial material is accompanied by open access digital resources that are linked in this article. 
\end{abstract}

\ccsdesc[500]{Hardware}
\ccsdesc[500]{Security and privacy~Hardware attacks and countermeasures}

\keywords{hardware, cybersecurity, scan chain, logic locking}

\maketitle

\section{Introduction}

Cybersecurity is a system-level challenge with various assets and threats at all levels of abstraction. 
Throughout the entire lifecycle of embedded computer systems, designers and end-users need to be aware of risks arising from untrusted parties~\cite{rostami_primer_2014}. 
In digital system design, we are usually concerned with meeting power, performance, and area objectives, where we might also try to improve testability, reliability, and other non-functional properties. 

In this tutorial, we present an introduction to basic concepts in the domain of \textbf{hardware security}. So that non-experts may gain insights into the mindset and challenges when working in this domain, we will also take you through a hands-on journey using two case studies\footnote{This tutorial is accompanied by a set of online materials, available at \url{https://github.com/learn-hardware-security}, which you can use to follow the case studies.} from domains which currently have active research communities.

\textbf{Why become more familiar with hardware security?} The semiconductor industry continues to become more valuable (possibly reaching US\$600 billion in 2022~\cite{deloitte_2022_2022}), with chips used in multiple end markets. Researchers and practitioners in digital design need to be well-equipped to deal with threats that arise whenever one produces something of value. 
Industry and governments continue to invest heavily in tools and techniques for improving hardware security. 
For example, at the time of writing, the United States' Defense Advanced Research Projects Agency (DARPA) funds the Automatic Implementation of Secure Silicon (AISS) program, which tries to ``ease the burden of developing secure chips... [and] provide a means of rapidly evaluating architectural alternatives that best address the required design and security metrics, as well as varying cost models to optimize the economics versus security trade-off''~\cite{darpa_darpa_2020}. The Semiconductor Research Corporation (SRC), an industry consortium, features hardware security as one of its priority areas~\cite{semiconductor_research_corporation_semiconductor_2022}, launched following a National Science Foundation (NSF) joint workshop~\cite{computing_community_consortium_research_2013}. The European Union is pushing for stronger standards in cybersecurity in both hardware and software~\cite{european_commission_state_2022}, which will require more expertise in security-aware design. These are only a few examples of the strong interest in this fast-moving domain, and we hope readers of this tutorial will become inspired to make new contributions to hardware security. 
Through the case studies in this tutorial, we hope that you will gain an appreciation of how hardware mechanisms need to be scrutinized for their impact on security, as well as gain an understanding of how hardware intellectual property could be protected. 
In carefully considering the motivations and processes in these case studies, you should become better equipped to explore other areas of hardware security.

The tutorial is laid out as follows.
In \autoref{sec:principles}, we will talk about hardware security in the broadest sense.  
The first case study is presented in \autoref{sec:scan}, where we will take you through an end-to-end attack on a commonly utilized technique for post-fabrication testing of digital circuits: i.e., an attack on \emph{scan chains}. We will show how an embedded cryptographic key is vulnerable to leaks, even when the key is not itself exposed to the scan chain. 
Then, in \autoref{sec:locking}, we will take you through the basics of logic locking from the ground up. Logic locking is a set of evolving techniques focused on the protection of hardware intellectual property from reverse engineering and piracy, particularly in the context of an untrusted foundry. This case study will work through the motivations and foundations of logic locking and then provide a step-by-step walk-through of one of the pivotal attacks in the literature: the SAT attack~\cite{subramanyan_evaluating_2015}. 

We assume that you, the reader, have a basic understanding of the fundamental topics related to digital system design (e.g., digital logic gates, Boolean algebra) but \textbf{no prior knowledge about scan chain security or logic locking}; thus we will start by making you more familiar with security generally before a hands-on dive into those topics. 
Readers with a bit more background will find the material a good refresher and we encourage you to sample some of the more recent work mentioned throughout the tutorial and at the end of each case study. 

After working through this tutorial, you should: 
\begin{itemize}
    \item be able to apply a \textbf{security mindset} in thinking about a system;
    \item understand the role of the scan chain, its benefits, and risks to the system;
    \item be able to work through a scan-based exploit of a cryptographic design;
    \item understand the motivations for logic locking, its overall objectives, and the basic principles of the area; and
    \item be able to analyze and think critically about simple locking. 
\end{itemize}

\section{Principles of Hardware Security \label{sec:principles}}

\subsection{The Broader Context}

All trustworthy software functionality fundamentally relies on the correct operation of trustworthy underlying hardware.
For instance, when we load our banking data via our web browsers, we assume that the computer powering the web browser does not surreptitiously save our details or leak them to third parties.
When we use our bank card to make a purchase, we assume that the hardware doesn't carelessly leave our details in memory.
Yet, hardware is often designed with only the desired functionality in mind, not essential security properties, and there are numerous examples in the literature of hardware being broken (e.g. smart cards~\cite{markantonakis_attacking_2009}, microcontroller memory protection fuses~\cite{strobel_microcontrollers_2014}, and on DRAM memory systems such that they leak sensitive data~\cite{mutlu_rowhammer_2020}).

While functionality is of course important, as are other design metrics (cost, power consumption, performance, size, and reliability among them), leaving security as an afterthought is dangerous, and the increasing number of hardware-based attacks is highlighting this~\cite{rostami_primer_2014,basu_cad-base_2019}. 
That said, industry has recently started acknowledging these risks, and initiatives such as the MITRE Hardware CWEs~\cite{the_mitre_corporation_mitre_cwe_2022,the_mitre_corporation_cwe_2022} have begun to classify known design weaknesses. 

One major driver of industry adoption comes from the risks inherent in integrated circuit (IC) and printed circuit board (PCB) counterfeiting and reverse engineering. Electronic supply chains are distributed worldwide~\cite{rostami_primer_2014}, introducing many possible points of attack.
Designing an IC or PCB involves the creation of intellectual property (IP) that may come from third-party organizations or in-house or both, then integrating those components, and generating an IC/PCB layout; a blueprint of the design will then be sent to the manufacturer. Post manufacturing, the designs will be tested, which may be at yet another organization, before being packaged, distributed, and sold, again, with other parties in the loop. 
Given the high value represented by IP, there is considerable motivation to prevent reverse-engineering or piracy; for example, reverse-engineered ARM microcontrollers available on the grey market with reduced security~\cite{obermaier_one_2020} represent supply-chain risks. 

Further compounding the security challenge are the risks of faults entering into the design, either accidentally or maliciously added. Hardware Trojans~\cite{xiao_hardware_2016} refer to the category of maliciously introduced fault-inducing artifacts, and they can be caused by adding, altering, or removing components or connections.
Unfortunately, designing a product such that it can be easily tested can also provide attack vectors for malicious third parties. 

\textbf{So how do we begin to get a handle on hardware security?}
In this tutorial, we will use as pedagogical examples two case studies to introduce you to hardware security concepts. 

The first is on attacking systems via exploiting scan chains (\autoref{sec:scan}), which illustrates how ``helpful'' hardware components can be manipulated for nefarious purposes. 
This example serves to demonstrate hardware-based exploitation, thus motivating research work on secure scan and other hardware-assisted security approaches. 
The second provides an intuitive introduction to logic locking (\autoref{sec:locking}), culminating in a walk-through example application of the pivotal ``SAT attack''~\cite{subramanyan_evaluating_2015}. 
This example serves as an on-boarding into the domain of intellectual property protection, providing insights into the motivations that drive the ``cat-and-mouse'' of protecting hardware \textit{itself} from reverse-engineering analysis, theft, and so on. 
Both case studies focus on the \textit{confidentiality} property of security, being the protection of cryptographic keys (in the scan chain example) and the functionality of a design (in the logic locking example). 
Readers who are already familiar with security properties should feel welcome to proceed to \autoref{sec:scan}; otherwise, the next section provides a quick introduction to security as framed by ``the CIA triad.''

\subsection{Confidentiality, Integrity, and Availability - the CIA Triad}
\label{sec:cia-triad}

The CIA Triad of confidentiality, integrity, and availability defines the central tenants of all cybersecurity~\cite{samonas_cia_2014, chapple_confidentiality_2018}. 
All attacks and defenses can be viewed within the context of the Triad, and for any device or product to be considered secure, it must address all three facets. These three properties can apply at various levels of abstraction, from the hardware design level through to the system level (potentially involving distributed systems). 

\textbf{Confidentiality} refers to the protection of data and resources from unauthorized access. For example, your credit card stores important banking details in its embedded ICs. These details need to be well-protected to prevent third parties from being able to copy them such that they may use them fraudulently to create new purchases of their own. There are many countermeasures to protect confidentiality, including password access and other access controls, encryption, and physical defenses.

\textbf{Integrity} covers the defense of information from unauthorized alteration. For instance, it should not be possible to alter your transaction (e.g. change the dollar value), once confirmed by the electronics in your bank card.
Integrity-based measures provide assurance that data is both accurate and complete. In electronic systems, it is not only necessary to control access to the components in the digital realm, but also in the physical realm, and ensure that authorized users can only alter the information that they are legitimately authorized to alter. Examples of countermeasures in this space can also be based on encryption, such as taking digital hashes or signatures of files; or by duplicating data and storing it in multiple ways.

\textbf{Availability} refers to the need for a given system to be accessible by authorized users. Here, malicious attacks seek to prevent this access and so are often referred to as Denial of Service (DoS). Examples include cryptolockers and ransomware - for instance, the 2021 Colonial Pipeline attack which impacted the availability of gas on the east coast of the United States~\cite{sanger_pipeline_2021}. 
Within the context of a hardware-based system, an availability-based attack example could be as simple as stealing the aforementioned banking card or damaging the reader. 
Availability countermeasures can involve hardware and software duplication and redundancy (i.e., backup systems), and special types of hardware and software to detect if an attack is occurring.

\emphbox{\textbf{Insight:} Given your own experience, reflect on things you consider valuable in a given system. These things are \textit{assets}. Imagine what could go wrong if any of the three properties of Confidentiality, Integrity, or Availability are violated.}

\subsection{Hardware Security Domains}
There are many areas that fall under the umbrella of \textit{hardware security}. 
For instance, Rostami \textit{et al.}~\cite{rostami_primer_2014}\footnote{We reference a number of papers in this section as potential starting points for further reading, but please note that there are many, many works out there---our list is not exhaustive. Any apparent omissions are unintentional, and in fact, we challenge the reader to find the literature that is relevant to your interests, beyond what we've mentioned here. Happy hunting!} pose several hardware security problems, such as IP piracy, reverse engineering, side-channel analysis, counterfeiting, and malicious implants (like backdoors or hardware Trojans) each with a vibrant community of researchers and practitioners exploring new attacks and defenses. 
Hardware security research also considers how the security of an overall system can be enhanced or mitigated by hardware, such as by the addition of security mechanisms (e.g., dedicated security chips~\cite{kostiainen_dedicated_2020}) or by attacks enabled by the oversights in a hardware design (e.g., Hermes (PCIe)~\cite{zhu_hermes_2021}, Spectre~\cite{kocher_spectre_2019}, Meltdown~\cite{lipp_meltdown_2018}, RowHammer~\cite{kim_flipping_2014}, Scan-chains~\cite{yang_scan_2004}).
From the embedded systems domain~\cite{ravi_security_2004} through to the cloud~\cite{coppolino_comprehensive_2019}, hardware security is crucial. 

\textbf{How can one digest such a wide field?} One strategy is to consider hardware security as comprising \textit{hardware for security}, i.e., hardware-assisted security (where surveys such as that by Tan~\cite{tan_challenges_2022}, Jin~\cite{jin_towards_2019}, Coppolino \textit{et al.}~\cite{coppolino_comprehensive_2019}, and Brasser \textit{et al.}~\cite{brasser_special_2018} might be useful starting points), and \textit{security for hardware}, i.e., the area of protecting hardware designs themselves (for this area, works such as that by Rostami \textit{et al.}~\cite{rostami_primer_2014}, Chakraborty \textit{et al.} ~\cite{chakraborty_keynote_2020}, and Xiao \textit{et al.}~\cite{xiao_hardware_2016} could provide good starting points for further reading). 
In examining these works, one should appreciate that security is rarely considered the sole design objective and that there is a tension between security and traditional goals to reduce power consumption/area while increasing performance. 
It remains an open problem to characterize the trade-offs between security and other design metrics, partly because notions and metrics for security vary (e.g., in logic locking, different works evaluate security success and impacts differently~\cite{tan_benchmarking_2020}).

\emphbox{\textbf{Insight:} When you make your own way through the literature, keep an eye out for different elements, including, but not limited to: the motivation for the work (often, the threat model), the metrics used for evaluation (both in terms of security and its impacts on other design factors), and the experimental platforms used. Do you notice any commonalities in each area of hardware security?}

\section{Case Study: A tutorial on attacking scan chains \label{sec:scan}}

\subsection{Overview}
Scan chains, a Design  for Test (DFT) technique, are implemented in integrated circuits (ICs) in order to test their correct functionality~\cite{chang_functional_1998,huang_survey_2008}. They provide high fault coverage and do not need complex hardware for test pattern generation or signature analysis.

Fundamentally, a scan chain is a sequential combination of internal registers / flip-flops. 
They are constructed during synthesis by modifying normal D flip-flops to also include scan logic.
This scan logic consists of multiplexers which are daisy-chained so that the D flip-flops can be disconnected from the main combinational circuit to instead sequentially feed data between themselves. 

\begin{figure}[b]
    \centering
    \includegraphics{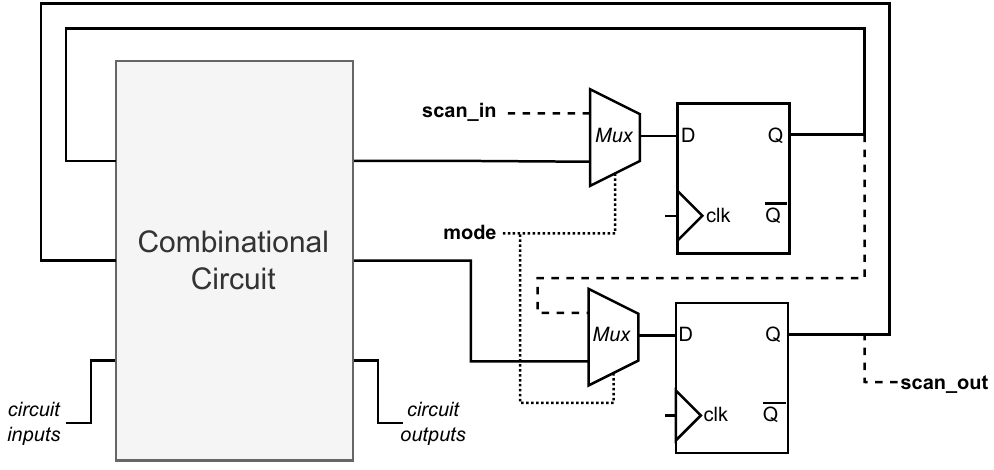}
    \caption{Generalized scan chain architecture}
    \label{fig:scan-chain-arch}
\end{figure}

The generic architecture of a scan chain is depicted in \autoref{fig:scan-chain-arch}. As shown, when the circuit is operating normally, the D flip-flops function normally, taking and returning state to the combinational circuit. However, when the circuit is put into test mode, each D flip-flop is joined using the multiplexers such that their contents may be serially `shifted' (as if they were a shift register) between one another and out of the circuit. In this context the shifting is referred to as scanning.

In typical implementations of scan chains, the hardware is connected to a five-pin serial JTAG~\cite{IEEE_IEEE_2013} boundary scan interface, where TCK is the test clock signal, TMS selects either normal mode or scan (test) mode, TRST is the reset control signal, TDI is connected to the input of the scan chain and is used to scan in new values, and TDO is connected to the output of the scan chain and is where the internal register values will appear.

However, scan chains represent a significant source of information about the internal functionality and implementation of a given device. As such, designers will typically seek to protect access to the scan architecture. The simplest method for this is to leave access to the scan chain unbound when physically packaging the IC. However, as packages can be broken open for reverse engineering purposes, this is surmountable. Alternatively, or in addition to, the access to the scan chain or JTAG interface can be controlled via other methods, including setting protection bits, fuses, or using access control passwords. However, given enough time with physical access to the component, protections such as these can still be compromised.

In this tutorial case study, we will examine how a compromised scan chain can be utilized to leak secrets about a given electronic circuit. 
We consider the case where a cryptographic algorithm is implemented as an application-specific integrated circuit (ASIC). Specifically, we consider the symmetric-key Data Encryption Standard (DES)~\cite{nist_data_1999}. 
We choose DES in this tutorial for three reasons: (1) It is a relatively straightforward algorithm that we can comprehensively describe within this tutorial, ensuring a self-contained case study; (2) Although retired (no longer recommended for new applications), DES was a widely-adopted algorithm used for decades to protect digital secrets (and still sees use within legacy systems); and (3) the process that will be illustrated in this tutorial is also suitable for attacking other encryption standards, including the more advanced and current best-practice Advanced Encryption Standard (AES)~\cite{nist_advanced_2001} algorithm. 
Scan chain attacks on both DES~\cite{yang_scan_2004} and AES~\cite{ali_test-mode-only_2014,yang_secure_2006} have both been previously demonstrated in the literature.

\emphbox{\textbf{Insight:} 
How does this example fit into the CIA Triad (\autoref{sec:cia-triad})?
A digital circuit with an embedded encryption key would be expected to keep this key \emph{confidential}. 
For example, consider how an encrypted digital media stream (e.g. satellite TV) should only be broadcast to authorized users (e.g. those with the decoding hardware). In order to maintain this ideal, the decoding hardware should ensure that the embedded cryptographic keys are well-protected. 
}

\subsection{Data Encryption Standard (DES)}

\begin{figure}
    \centering
    \includegraphics[width=0.85\textwidth]{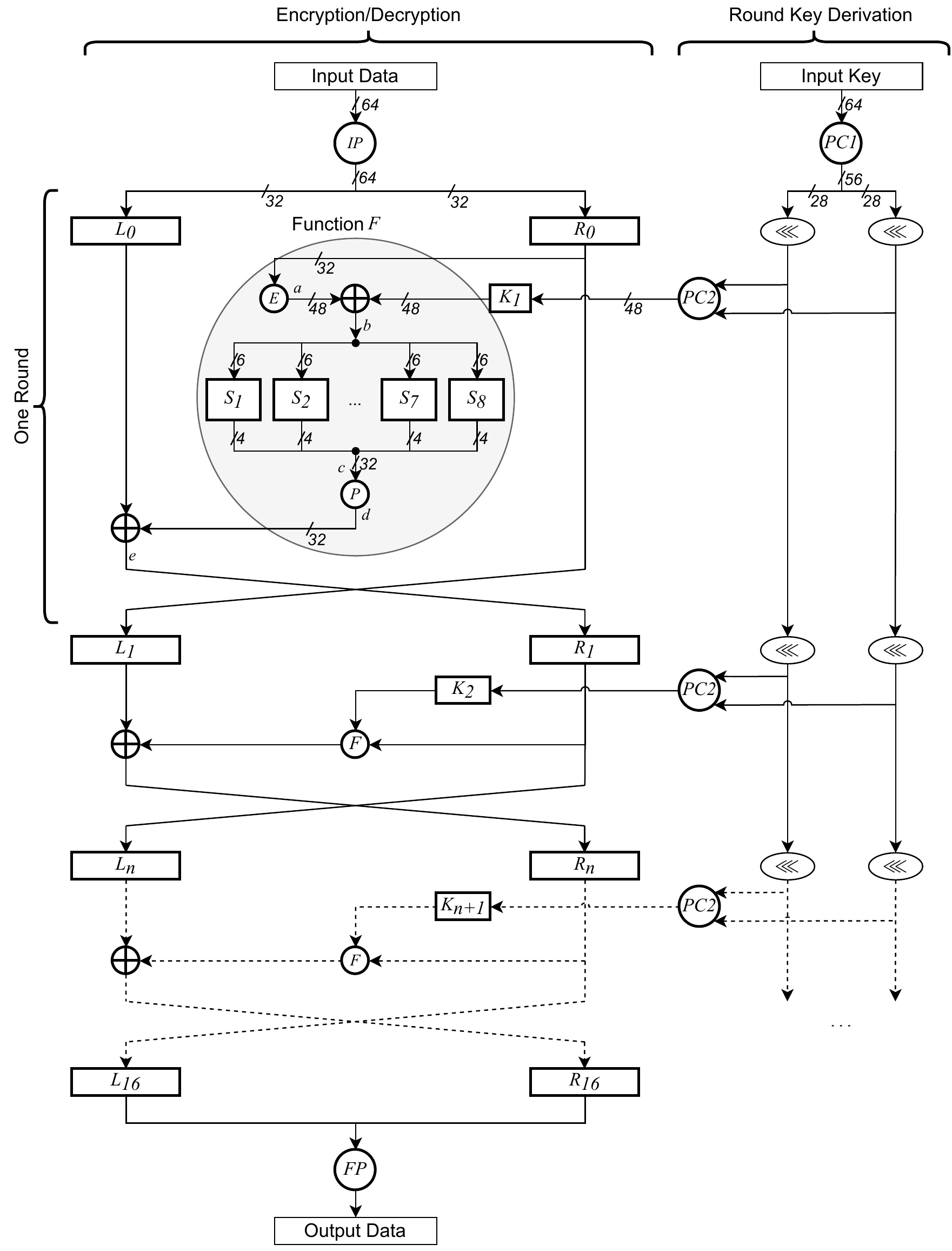}
    \caption{High level structure of DES.}
    \label{fig:des-high-level}
\end{figure}

The Data Encryption Standard (DES) is a symmetric-key block cipher published by the National Institute of Standards and Technology (NIST) in the year 1977, and most recently updated in 1999~\cite{nist_data_1999}. It encrypts 64 bits of data at a time, using a 56-bit key (usually stored as a 64-bit value with checksum bits). 
DES is a Feistel Cipher implementation, meaning it utilizes a repeating block structure and both encryption and decryption utilize the same algorithm.
DES is based on 16 rounds of 64-bit blocks. The high-level structure of the algorithm is presented in \autoref{fig:des-high-level}. 
It is separated into two parts, \emph{round key derivation} and \emph{encryption/decryption}. 

\textbf{Round Key Derivation:}
This is depicted in the right-hand side of \autoref{fig:des-high-level}.
Here, we input the 64-bit key and perform a permutation using table PC1 which re-orders the bits (see \autoref{tbl:des-IP} in \autoref{sec:appendix:des-tables}). PC1 both removes the superfluous 8 parity bits and splits and reorders the remaining 56-bits of key value into two 28-bit halves. 
Within each round, these two halves are left-rotated independently, either 1 or 2 bits, depending on the specific round of operation (see \autoref{tbl:des-SHIFTS} in \autoref{sec:appendix:des-tables}).
These two halves are then concatenated and \emph{compressed} to 48-bits by using the PC2 permutation table (see \autoref{tbl:des-PC2} in \autoref{sec:appendix:des-tables}).
This output is termed as a \emph{round key}.
Because the PC2 inputs are rotated between each round, the output of PC2 will change each round---each round will have a different round key.
Given that round key derivation is entirely separate from data encryption and decryption, it is possible to precompute and store the round keys if the intended encryption/decryption key is static.

\emphbox{\textbf{Insight:} 
There are 16 round keys, each 48 bits in size. However, we do not need to break $2^{(16*48)}$ bits of encryption.
The round keys were originally derived from the same key, and although this key was 64 bits originally, 8 of the bits are parity bits, and so there is effectively only 56 bits of entropy protecting the encryption. 

The 56-bit key length is what makes DES unsuitable for protecting secrets in the modern era---it is short enough that it can be brute forced, and has been vulnerable to this for many years (for example, see the EFF DES Cracker from 1998~\cite{van_de_zande_day_2001}). 
}

\textbf{Encryption/decryption:}
This is depicted in the left-hand side of \autoref{fig:des-high-level}.
Here, we input the 64-bit data.
This is first permuted using the Initial Permutation (IP) table (see \autoref{tbl:des-IP} in \autoref{sec:appendix:des-tables}) before the left-hand and right-hand 32-bits are separated into $L_0$ and $R_0$.

Then, in each round $n$ of the encryption / decryption process, the 32-bit value $R_n$ is taken and combined in function $F$ with the appropriate round key $k_{n+1}$.
When encrypting, the round keys are used in the forward order (as is shown in \autoref{fig:des-high-level}). When decrypting, the round key order is reversed, but  no other changes need to be made to the process. 

Function $F$ in \autoref{fig:des-high-level} is detailed as follows.
Firstly, the input from the $R$ register is \emph{expanded} using expansion permutation function $E$ (see \autoref{tbl:des-E} in \autoref{sec:appendix:des-tables}) to 48-bits, making value $a$. This is combined with the appropriate round key $k_{n+1}$ using an exclusive-or operation to make value $b$.
Each consecutive six-bit block of $b$ is then passed through the appropriate substitution box $S$ (see \autoref{tbl:des-sbox-all} in \autoref{sec:appendix:des-tables}) which makes 4-bit blocks then concatenated to make value $c$.
Value $c$ is then passed through permutation $P$ (see \autoref{tbl:des-P} in \autoref{sec:appendix:des-tables}) to make value $d$, which is finally then combined using exclusive-or with the  the $L_n$ to make value $e$.
In the next round, $R_{n+1}$ takes the value $e$, and $L_{n+1}$ takes the previous $R_n$, unless it is the last round. In this case ($n=16$), the swap does not occur; and instead, the resultant value is concatenated, before being passed into the Final Permutation (FP) (see \autoref{tbl:des-FP} in \autoref{sec:appendix:des-tables}).
The final permutation is derived by using the inverse of the IP.
This results in the algorithm's final encrypted output.

\subsection{Attacking a practical DES implementation}

\begin{figure}
    \centering
    \includegraphics[scale=0.9]{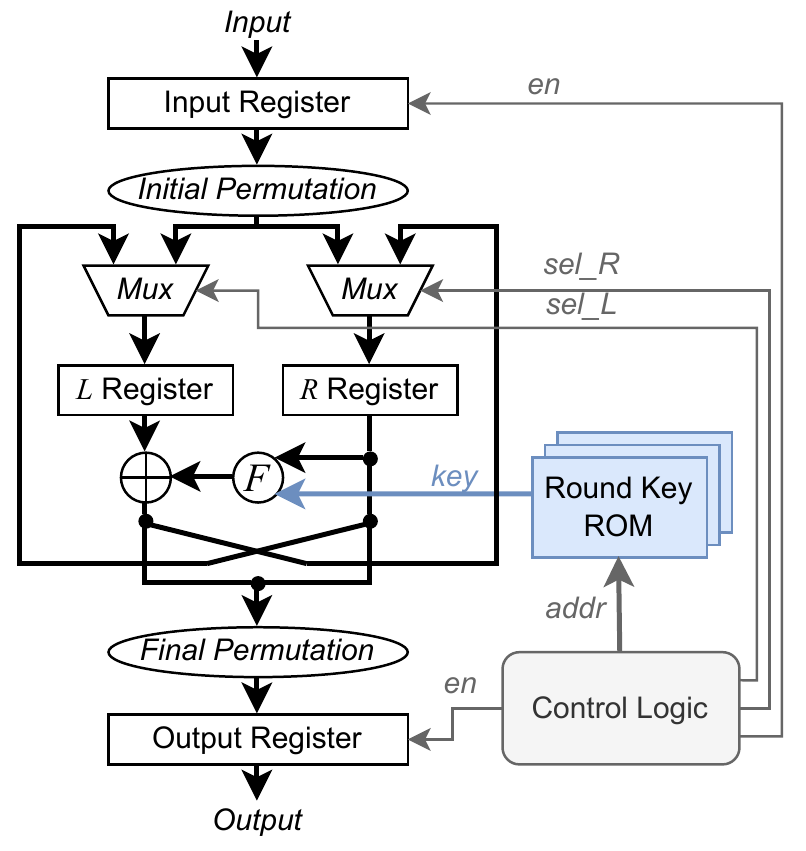}
    \caption{DES hardware block diagram}
    \label{fig:des-hardware}
\end{figure}

The DES implementation can be realized in hardware in a number of different ways.  One resource-efficient approach is to utilize a single set of L and R registers, and use these iteratively to perform each round of an encryption or decryption.
These can be combined with pre-computed round-keys stored in an on-chip read-only memory (ROM).
This approach is presented in \autoref{fig:des-hardware}. 
Internally, the initial permutation, final permutation, expansion E and permutation P functions are implemented as fixed one-to-one mappings. S-boxes are implemented either as gates or as ROMs. The DES controller primarily consists of a 4-bit counter which will index the appropriate round key for each round, as well as decode logic that will control the enable and select lines.

An implementation of this algorithm featuring a scan chain is what we focus on attacking, although the scan chain will not include the round keys ROM (otherwise the attack becomes trivial) or any registers from the control logic (we discuss this potential later, in \autoref{sec:design-explore}).

\textbf{What does an attacker know?} Firstly, we assume the attacker knows that the circuit they wish to attack is implementing DES. In addition, as it is public, the attacker knows the details of the DES algorithm. 
Secondly, the attacker will know certain details of the implementation.
For instance, the attacker would likely be able to obtain vendor-provided timing diagrams of the system. This is important, as from these, the attacker can infer the structure of the given DES circuit - for instance, whether or not it is iterative, pipelined, or purely combinational. In this case study, our architecture is iterative, meaning it operates one round per clock cycle. In the first clock cycle, the permuted input will be loaded to the $L$ and $R$ registers. In the next 15 clock cycles, the $L$ and $R$ registers will be iteratively updated by the result of the round key computations. 
The output register will be loaded after the final $L$ and $R$ values are permuted. Likewise, the attacker can determine that the round keys are stored internally in the design (in a ROM). 

It is reasonable to assume that the attacker will gain access to the scan chains, either via a JTAG port or via breaking open the IC package and directly probing the scan chain ports. 
However, the attacker will not know the structure of the scan chain. These are typically determined according to the position of the register in the physical layout of the circuit, and would not be known unless the digital design files of the IC (a more protected form of IP than the typically vendor-provided architecture and timing diagrams) were obtained.

Based on the above, we thus need to break our attack into two phases.
The first will determine the structure of the scan chain. 
The second will retrieve a DES round key, and from this, obtain the original DES key. 
Note that while we follow similar steps to that presented in Yang et. al's work~\cite{yang_scan_2004}, we use an alternative, more consistent nomenclature (where the `first bit' bit 1 will always refer to the left-most or most significant bit of a value or register) with alternative constants, present the attack steps in more detail (including accompanying code and code discussion), and use an alternative methodology for determining the final key. %

\emphbox{\textbf{Insight:} 
In this subsection we detailed what the hypothetical attacker does and does not know. 
This process is more formally known as \emph{attacker} or \emph{threat modeling}~\cite{shostack_threat_2014}, and it is an important step when considering attacks and defenses for given cybersecurity threats.
}

\subsubsection{Following the case study}
\label{sec:following}
In the rest of this section, we proceed with examples and code written in the Python programming language.
We use Python version 3.8.10.
The listings are designed to be run sequentially and in-order, and are provided with example outputs in immediately-following listings.
E.g., Copying and running Listing 1 will give the output of Listing 2, then copy and run Listing 3 to get Listing 4, and so on.
Most listings terminate with an \texttt{exit()} command (and possibly some \texttt{print} statements before these). You can comment these out safely, they are presented only for pedagogical purposes.
\textbf{The complete code for this case study, as well as the accompanying DES implementation, are available in our open repository at \url{https://github.com/learn-hardware-security/py-des-scan}}.
This also includes links to a Google Colab/Jupyter notebook.

\subsection{DES implementation (Python)}

\begin{figure}[h]
    \centering
    \includegraphics{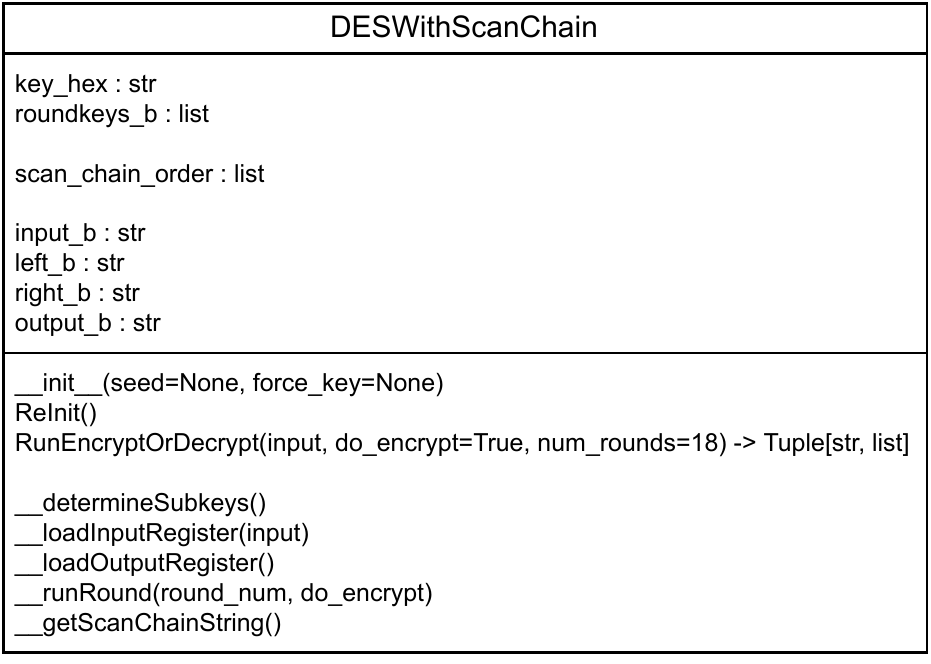}
    \caption{Simplified UML diagram for \texttt{DESWithScanChain} Python Class (attributes and public/private methods).}
    \label{fig:python-des-diagram}
\end{figure}

We can emulate the DES implementation from \autoref{fig:des-hardware} according to the Python class illustrated in \autoref{fig:python-des-diagram}. 
We define it as having only three public methods (assume all attributes are private). 
\begin{itemize}
    \item \texttt{\_\_init\_\_(seed=None, force\_key=None)}, which is used to instantiate the DES implementation. A random seed should be provided to derive the random structure of the scan chain, and if no key is provided in \texttt{force\_key}, will also randomly generate a key. If no seed is provided, the program will set it to numeric 1. In this tutorial, we will focus on determining the value of the key when one was not provided.
    \item \texttt{ReInit()} is used to reset the hardware between runs. It does not recompute keys.
    \item \texttt{RunEncryptOrDecrypt(input, do\_encrypt, num\_rounds) -> Tuple[str, list]} is the all-in-one function used to run the emulated hardware. It takes a 16-character hexadecimal string as \texttt{input}, as well as a boolean \texttt{do\_encrypt} that determines if the hardware should encrypt (\texttt{True}) or decrypt (\texttt{False}). It also takes a number of execution rounds \texttt{num\_rounds}, which refers to the number of clock cycles the hardware should be provided when executing. 
    
    The function returns two values as a Tuple.
    The first of these is the value of the output register, encoded as a 16-character hexadecimal string. The second of these is a list of scan chain outputs, as observed between every \emph{round} of execution (i.e. every tick of the encryption hardware).
    The length of this list will thus be the same as the provided \texttt{num\_rounds} argument.
\end{itemize}

We now instantiate the design according to the code presented in \autoref{lst:des-dut-instantiate}. 

\textbf{Running example}. In this case study we will execute the code at each stage of the analysis. For the example DES instantiation provided, the output is presented in \autoref{lst:des-dut-instantiate-output}.
For a more interactive presentation, consider obtaining the code from the associated GitHub repository (see \autoref{sec:following}).

\begin{lstlisting}[style=PythonStyle,caption={Tutorial Setup and Instantiating the DES Design Under Test (DUT)},label={lst:des-dut-instantiate}]
#Useful modules for this tutorial
from typing import *
import itertools

#Import the DESWithScanChain module 
# assuming the py-des-scan GitHub repository is downloaded to a folder called 'py_des_scan'
import py_des_scan.des_scan as des

#Define a random seed for the emulated hardware's key and scan chain
seed = 6

#Instantiate the DES module that we will test/attack
dut = des.DESWithScanChain(seed)

#Do a test run of the DES with a given input 
test_code = "0BADC0DEDEADC0DE"
print("Input: " + test_code)
(check_ciphertext, _) = dut.RunEncryptOrDecrypt(test_code)
print("Ciphertext: " + check_ciphertext)
(plaintext, _) = dut.RunEncryptOrDecrypt(check_ciphertext, do_encrypt=False)
print("Plaintext: " + plaintext)
\end{lstlisting}

\begin{lstlisting}[style=TextStyle,caption={Example Output for Listing~\ref{lst:des-dut-instantiate}},label={lst:des-dut-instantiate-output}]
Input: 0BADC0DEDEADC0DE
Ciphertext: 5FB5CD14D3136003
Plaintext: 0BADC0DEDEADC0DE
\end{lstlisting}

\subsection{Attack Phase 1: Determining the Scan Chain Structure}

The first part of the attack deals with reverse-engineering the locations of the flip-flops for the Input, L, and R registers in the scan chain.
This is necessary as the scan chain output does not intrinsically reveal the correspondence between the data values it provides and the registers internal to the design. 
The general process for this is follows.
\begin{enumerate}
    \item Reset the DES hardware to clear all registers.
    \item Present DES hardware with input $(8000000000000000)_{16}$ (i.e. 64-bits with the left-most bit (i.e. MSB) set).
    \item Run it for one clock cycle such that the input register is loaded.
    \item Scan out the bit stream pattern (Pattern 1).
    \item Pattern 1 will have one bit active. This position corresponds with the position of the input's currently active bit (i.e., the MSB).
    \item Run it for one additional clock cycle so that the input is loaded into the L and R registers (after they pass through the initial permutation step). 
    \item Scan out the bit stream pattern (Pattern 2).
    \item Pattern 2 will have two bits active. As the input is not cleared after loading L and R registers, one of these will match the bit in Pattern 1 and can be disregarded. The other bit represents the bit position in the L or R register after passing through the initial permutation (for example, the first bit of the input will pass through to the 8th bit of the R register).
    \item Repeat steps 1-8, shifting the input by 1 bit each time, 63 more times to determine the position of the remaining input and L/R register bits.
\end{enumerate}
Python code to complete this is presented in \autoref{lst:scan-chain-reverse}.

\begin{lstlisting}[style=PythonStyle,caption={Code to determine Input, L, and R positions in scan chain},label={lst:scan-chain-reverse}]
#Define arrays for storing the determined indices in the random scan chain
input_scan_indices = [None] * 64
left_r_scan_indices = [None] * 32
right_r_scan_indices = [None] * 32

#Input a single bit in each of the 64 possible positions and run two rounds, 
# capturing the scan chains of each cycle.
# In the first cycle, we can determine that bit of the input register
# In the second cycle, we can determine the bit of the L/R register
for i in range(64):
    dut.ReInit() #Reset the hardware 
    
    #Determine the input hex string
    input_num = 1 << (63 - i) 
    input_hexstr = '%016X' % (input_num)
    
    #Get scans for the input (we run in Encrypt mode)
    (_, scans) = dut.RunEncryptOrDecrypt(input_hexstr, True, 2)
    
    #The scans are 192 bits (represented as ASCII 0/1 characters) long. 
    # In scans[0], only one bit will be True; 
    # this represents the i-th bit in the Input register.
    for j in range(192):
        if scans[0][j] == "1":
            input_index = j
            break
    #We can store this immediately, using "i" as the position
    input_scan_indices[i] = input_index;
    
    # In scans[1], two bits will be True;
    # the one not present in the first scan represents the i-th bit in the L/R registers 
    # after Initial Permutation.
    for j in range(192):
        if scans[1][j] == "1" and j != input_index:
            lr_index = j
            break
    #We need to invert the initial permuatation before we can store this
    # For this we can use the Final Permuation table as this is the pre-computed inverse
    lr_pos = des.FINAL_PERM[i] - 1; #table values are 1-indexed
    #The low 32 lr_pos values refer to the L register, high values to R register
    if lr_pos < 32:
        left_r_scan_indices[lr_pos] = lr_index
    else:
        right_r_scan_indices[lr_pos - 32] = lr_index

# Example output
print("input_scan_indices:", input_scan_indices)
print()
print("left_r_scan_indices:", left_r_scan_indices)
print()
print("right_r_scan_indices:", right_r_scan_indices)
\end{lstlisting}

\begin{lstlisting}[style=TextStyle,caption={Example Output for Listing~\ref{lst:scan-chain-reverse}},label={lst:scan-chain-reverse-output}]
input_scan_indices: [20, 152, 61, 26, 71, 78, 145, 110, 60, 36, 11, 2, 176, 140, 85, 130, 55, 32, 111, 74, 179, 106, 27, 17, 83, 129, 79, 92, 182, 43, 125, 180, 141, 42, 49, 103, 167, 191, 40, 95, 12, 155, 146, 21, 188, 168, 153, 7, 166, 147, 3, 75, 173, 161, 33, 119, 70, 0, 171, 35, 144, 73, 25, 139]

left_r_scan_indices: [156, 80, 118, 142, 186, 151, 131, 160, 162, 123, 45, 58, 124, 23, 184, 127, 10, 117, 143, 63, 189, 190, 159, 38, 99, 102, 51, 132, 4, 47, 112, 15]

right_r_scan_indices: [13, 116, 62, 177, 66, 72, 157, 54, 90, 50, 137, 31, 128, 28, 57, 170, 59, 1, 29, 91, 67, 172, 136, 134, 165, 185, 121, 120, 133, 37, 164, 34]
\end{lstlisting}

\subsection{Attack Phase 2: Determining Round Key 1}

Now that we know the location of the L and R values in the scan chain, we can load out their contents. A function to do this process is presented in \autoref{lst:scan-chain-read-l-r}.

\begin{lstlisting}[style=PythonStyle,caption={Code to read contents of L and R register from scan chain},label={lst:scan-chain-read-l-r}]
#Given the set of indices for the L and R registers and the given scan chain output,
# return the contents of the L and R registers.
def read_scan_l_r(left_scan_indices, right_scan_indices, scan) -> Tuple[list, list]:
    l_reg = [None]*32
    r_reg = [None]*32
    for i in range(32):
        l_reg[i] = scan[left_scan_indices[i]]
        r_reg[i] = scan[right_scan_indices[i]]
    return (l_reg, r_reg)

#Example read from l/r regs
test_code = "0BADC0DEDEADC0DE"
(_, scans) = dut.RunEncryptOrDecrypt(test_code, True, 2) 
(l_reg, r_reg) = read_scan_l_r(left_r_scan_indices, right_r_scan_indices, scans[1])
print("l_reg contains:", l_reg)
print()
print("r_reg contains:", r_reg)
print()    
\end{lstlisting}

\begin{lstlisting}[style=TextStyle,caption={Example Output for Listing~\ref{lst:scan-chain-read-l-r}},label={lst:scan-chain-read-l-r-output}]
l_reg contains: ['1', '1', '0', '1', '1', '1', '0', '0', '1', '0', '0', '1', '1', '0', '0', '0', '1', '0', '1', '1', '1', '0', '1', '0', '0', '0', '1', '0', '0', '0', '1', '1']

r_reg contains: ['1', '1', '1', '1', '1', '1', '1', '0', '0', '0', '1', '0', '0', '0', '1', '0', '1', '0', '1', '1', '1', '0', '1', '1', '1', '0', '0', '1', '1', '0', '0', '1']

\end{lstlisting}
We wish to observe the intermediate (less protected) encryption data.
We are specifically interested in the result after the first round of encryption.
Here, only the first round key $K_1$ has been applied to the data.
Consider \autoref{fig:des-high-level}.
Here, the first round of DES can be described using Equations~\ref{eqn:a}-\ref{eqn:l1}.
\begin{equation}\label{eqn:a}
    a = E(R_0)
\end{equation}
\begin{equation}\label{eqn:b}
    b = a \oplus K_1
\end{equation}
\begin{equation}\label{eqn:c}
    c = SBoxes(b)
\end{equation}
\begin{equation}\label{eqn:d}
    d = P(c)
\end{equation}
\begin{equation}\label{eqn:r1}
    R_1 = e = L_0 \oplus d
\end{equation}
\begin{equation}\label{eqn:l1}
    L_1 = R_0
\end{equation}

Consider the scenario where a known input is loaded into $L_0$ and $R_0$. 
An encryption round is run, generating $L_1$ and $R_1$, which are then scanned out using the scan chain.
Given that we know $R_0$, and $E$ is a simple permutation, we can derive $a$ using the inverse permutation $E^{-1}$ (\autoref{eqn:a}).
As we know $R_1$, we know $e$; and given that we also know $L_0$ we can compute $d$ (\autoref{eqn:r1}).
From this we can compute $c$ by taking the inverse permutation $P^{-1}$ (\autoref{eqn:d}).
All that remains is $b$, which we can then use to calculate $K_1$.

Let us explore the s-boxes further. 
In each round of DES, eight different s-boxes $\{S_1, S_2, \ldots, S_8\}$ are used.
The first, $S_1$, is presented in \autoref{tbl:des-sbox1}.
Six bits are used to determine which 4-bit value will be taken from an s-box (in other words, each s-box compresses 6 bits to 4 bits).
For $S_1$, it is the six left-most bits (most significant bits) $\{b_{1}, b_{2}, b_{3}, b_{4}, b_{5}, b_{6}\}$.
These are separated into row and column indices as indicated in the table.
The first and last bits $b_{1}$ and $b_{6}$ are used to determine the s-box row.
The middle four bits, $b_{2\ldots5}$, will be used to determine the s-box column.
For example, input $b_{1..6} = (100100)_2$ uniquely identifies value 14 from the table (row 3, column 3).

\begin{table}[h]
\caption{DES s-box $S_1$}
\label{tbl:des-sbox1}
\resizebox{\linewidth}{!}{%
\begin{tabular}{r|cccccccccccccccc|}
\cline{2-17}
\textbf{S1}                           & \multicolumn{1}{r}{\textbf{x0000x}} & \multicolumn{1}{r}{\textbf{x0001x}} & \multicolumn{1}{r}{\textbf{x0010x}} & \multicolumn{1}{r}{\textbf{x0011x}} & \multicolumn{1}{r}{\textbf{x0100x}} & \multicolumn{1}{r}{\textbf{x0101x}} & \multicolumn{1}{r}{\textbf{x0110x}} & \multicolumn{1}{r}{\textbf{x0111x}} & \multicolumn{1}{r}{\textbf{x1000x}} & \multicolumn{1}{r}{\textbf{x1001x}} & \multicolumn{1}{r}{\textbf{x1010x}} & \multicolumn{1}{r}{\textbf{x1011x}} & \multicolumn{1}{r}{\textbf{x1100x}} & \multicolumn{1}{r}{\textbf{x1101x}} & \multicolumn{1}{r}{\textbf{x1110x}} & \multicolumn{1}{r|}{\textbf{x1111x}} \\ \hline
\multicolumn{1}{|r|}{\textbf{0yyyy0}} & 14                                  & 4                                   & 13                                  & 1                                   & 2                                   & 15                                  & 11                                  & 8                                   & 3                                   & 10                                  & 6                                   & 12                                  & 5                                   & 9                                   & 0                                   & 7                                    \\
\multicolumn{1}{|r|}{\textbf{0yyyy1}} & 0                                   & 15                                  & 7                                   & 4                                   & 14                                  & 2                                   & 13                                  & 1                                   & 10                                  & 6                                   & 12                                  & 11                                  & 9                                   & 5                                   & 3                                   & 8                                    \\
\multicolumn{1}{|r|}{\textbf{1yyyy0}} & 4                                   & 1                                   & 14                                  & 8                                   & 13                                  & 6                                   & 2                                   & 11                                  & 15                                  & 12                                  & 9                                   & 7                                   & 3                                   & 10                                  & 5                                   & 0                                    \\
\multicolumn{1}{|r|}{\textbf{1yyyy1}} & 15                                  & 12                                  & 8                                   & 2                                   & 4                                   & 9                                   & 1                                   & 7                                   & 5                                   & 11                                  & 3                                   & 14                                  & 10                                  & 0                                   & 6                                   & 13                                   \\ \hline
\end{tabular}}
\end{table}

Each s-box outputs each possible value exactly four times. For instance, in s-box 1, the value $1$ is emitted for input $(000110)_2$, $(001111)_2$, $(100010)_2$, and $(101101)_2$.
Because of this, it is not possible to determine the input to an s-box by observing just one output.
However, each input to the s-boxes is the exclusive-or combination of the round-key bits as well as value $a$, which is computed from the expansion $E$ of the register $R$.
This means if we provide multiple values to be combined with the constant round key $K_1$ we can observe multiple different outputs based on the same key.
Given four possible outputs for each value, we thus need to use three different inputs.

Given we can set value $a$ arbitrarily, let us see how to reverse the key bits for the first s-box $S_1$.
Firstly, we apply our first input $a^1_{1..6} (000000)_2$.
Given \autoref{eqn:b} and \autoref{eqn:c}, that means we can derive \autoref{eqn:s1-c-1}.
\begin{equation}\label{eqn:s1-c-1}
    c^1_{1..6} = S_1(K_{1,1},K_{1,2},K_{1,3},K_{1,4},K_{1,5},K_{1,6})
\end{equation}
Given that each output of each s-box appears only once in each row, we shall switch one-bit in the column space (i.e. in the middle four bits of $b$). 
Thus, we apply our second input $a^2_{1..6} = (001000)_2$ to give our second output in \autoref{eqn:s1-c-2}.
\begin{equation}\label{eqn:s1-c-2}
    c^2_{1..6} = S_1(K_{1,1},K_{1,2},\overline{K_{1,3}},K_{1,4},K_{1,5},K_{1,6})
\end{equation}
We then switch two-bits in the input of the s-box $S_1$, changing both a row and a column. This is done through applying the third input $a^3_{1..6} = (010001)_2$, giving our third value \autoref{eqn:s1-c-3}.
\begin{equation}\label{eqn:s1-c-3}
    c^3_{1..6} = S_1(K_{1,1},\overline{K_{1,2}},K_{1,3},K_{1,4},K_{1,5},\overline{K_{1,6}})
\end{equation}
For each of Equations~\ref{eqn:s1-c-1}-\ref{eqn:s1-c-3} we will get four possible values for the key bits input to the s-box. However, only one of these values will be common to all three equations. This will be the final value of $K_{1,1...6}$. 

\begin{figure}[h]
    \centering
    \includegraphics[width=0.9\textwidth]{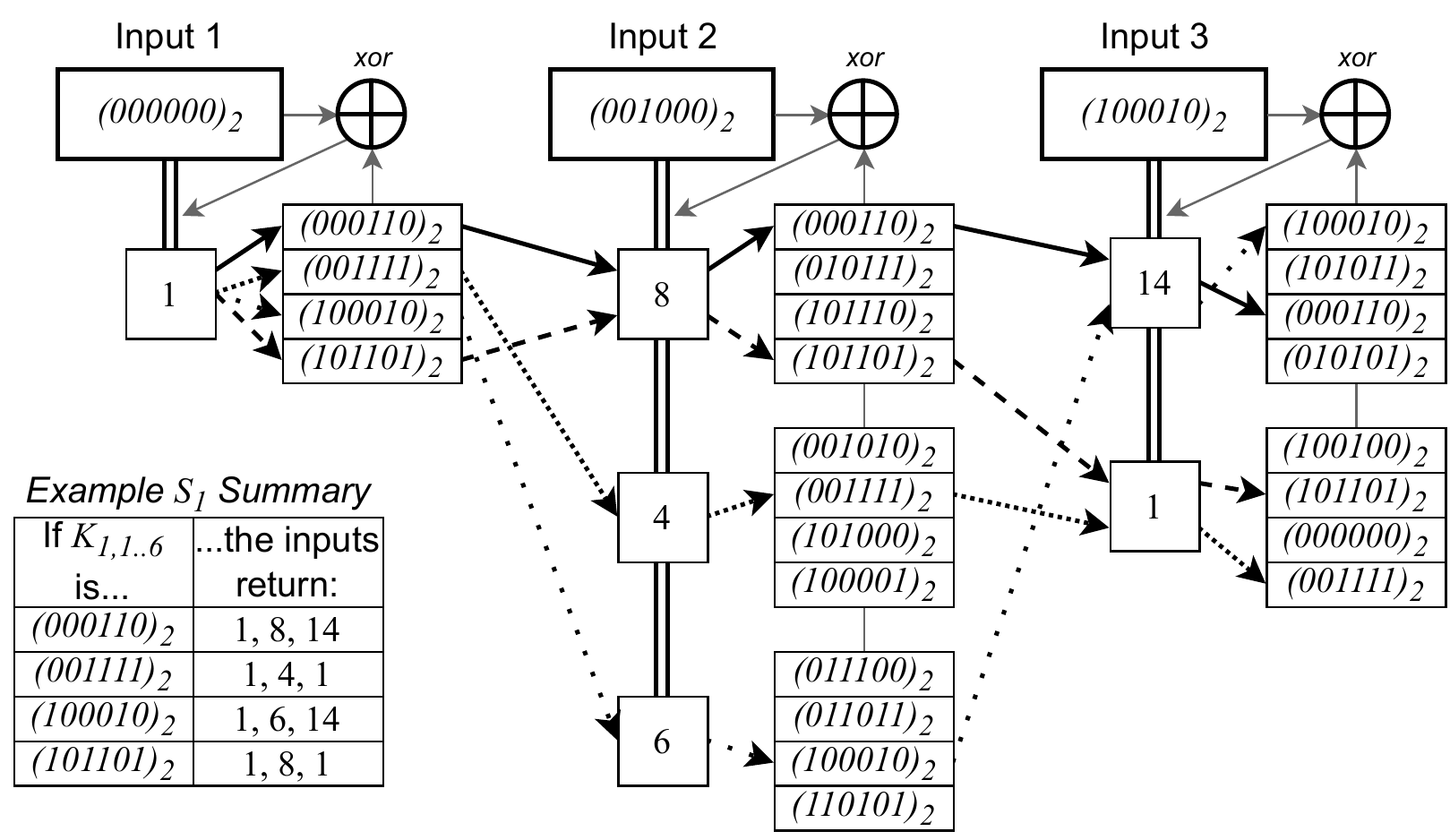}
    \caption{Reversing Round Key bits $K_{1,1..6}$.} %
    \label{fig:s1-reverse-when-1}
\end{figure}

Let us consider an example. We provide input $(000000)_2$ to point $a_{1..6}$, and observe that the s-box returns output $1$. Given that the input was all zeros, this means that the round key bits $K_{1,1..6}$ are one of $(000110)_2$, $(001111)_2$, $(100010)_2$, or $(101101)_2$.
We then provide input $(001000)_2$. If this returns value 8, then the only matching key bits that could emit this are $(000110)_2$ or $(101101)_2$.
If it returns value 4, then the only key bits that could do this are $(001111)_2$. Likewise, if it returns 6, the only possibility is that the key bits are $(100010)_2$.
If it did return 8, then we need to provide the third input, $(010001)_2$.
If this returns 14, then the only possibility is that the key bits are $(000110)_2$. Alternatively, if it returns 1, then the key bits must be $(101101)_2$.
This example is depicted pictorially in \autoref{fig:s1-reverse-when-1}, and we present the code for performing the s-box reversal in \autoref{lst:reverse-sbox}, the code also showing an example of how it can be used to obtain the values for s-box $S_1$.

\begin{lstlisting}[style=PythonStyle,caption={Code that reverses s-box outputs to a list of possible inputs},label={lst:reverse-sbox}]
#Given the concatenated output of the s-boxes (i.e. point 'c' in Fig. 2),
# (as a list of bits)
# and the concatenated input to the xor function (i.e. point 'a' in Fig. 2),
# (also as a list of bits)
# return the list of possible values for each s-box (i.e. list of lists).
def sboxes_output_to_possible_inputs(sboxes_output, sboxes_xor_input) -> List[list]:
    sboxes = []
    for i in range(8): #for each s-box
        #Get the output of _this_ s-box
        sbox_output = sboxes_output[i*4:(i+1)*4] 
        #Get the input to the xor for _this_ s-box
        sbox_xor_input = sboxes_xor_input[i*6:(i+1)*6]   
        
        #Convert the output of the s-box to an integer
        # (the des library stores s-box outputs as integers)
        sbox_value = 0
        for j in range(4): 
            sbox_value |= (int(sbox_output[j]) << (3 - j))

        #Find the 4 s-box inputs that produce the given output
        possible_sbox_inputs = []
        for row in range(4): #Every s-box value appears at least once in every row
            col = des.SBOXES[i][row].index(sbox_value)
            #print("Value %i Sbox %i row %i column %i" % (sbox_value, i, row, col))
            possible_input = [
                (row & 0b10) >> 1,
                (col & 0b1000) >> 3,
                (col & 0b100) >> 2,
                (col & 0b10) >> 1,
                col & 0b1,
                row & 0b1
            ]
            #for each bit, undo the XOR operation
            for k in range(len(possible_input)):
                possible_input[k] = possible_input[k] ^ sbox_xor_input[k]

            possible_sbox_inputs.append(possible_input)
        sboxes.append(possible_sbox_inputs)

    return sboxes
    
#Example s-box reversal for S1
print(sboxes_output_to_possible_inputs([0,0,0,1],[0,0,0,0,0,0]), end="\n\n")
print(sboxes_output_to_possible_inputs([1,0,0,0],[0,0,1,0,0,0]), end="\n\n")
print(sboxes_output_to_possible_inputs([1,1,1,0],[1,0,0,0,1,0]))
\end{lstlisting}
\vspace{-2mm}
\begin{lstlisting}[style=TextStyle,caption={Example Output for Listing~\ref{lst:reverse-sbox}},label={lst:reverse-sbox-output}]
[[[0, 0, 0, 1, 1, 0], [0, 0, 1, 1, 1, 1], [1, 0, 0, 0, 1, 0], [1, 0, 1, 1, 0, 1]]]

[[[0, 0, 0, 1, 1, 0], [0, 1, 0, 1, 1, 1], [1, 0, 1, 1, 1, 0], [1, 0, 1, 1, 0, 1]]]

[[[1, 0, 0, 0, 1, 0], [1, 0, 1, 0, 1, 1], [0, 0, 0, 1, 1, 0], [0, 1, 0, 1, 0, 1]]]
\end{lstlisting}
\vspace{-3mm}
If $c^1_{1..6}$ is a value other than 1, we can still use this method to identify the bits of $K_{1,1..6}$.
Likewise, we can construct similar patterns for the other s-boxes $S_2 \ldots S_8$. These patterns can then be combined to discover the round key $K_1$.
We derive these patterns here.

We first discuss how to prepare the input $a$. Recall from \autoref{fig:des-high-level} and \autoref{eqn:a} that it is constructed using expansion $E$ from the $R$ register.
The formula for this is presented as \autoref{eqn:a_from_e_r}. Note the constraints within the construction of $a$, that is, some bits need to match (e.g. $a_1$ and $a_{47}$ are both taken from $r_{32}$, so they must be the same---both either `1' or `0').
\begin{equation}\label{eqn:a_from_e_r}
\begin{split}
a_{1..48}=\{~~~~& \\
    & r_{32},r_{1},r_{2},r_{3},r_{4},r_{5}, \\
    & r_{4},r_{5},r_{6},r_{7},r_{8},r_{9}, \\
    & r_{8},r_{9},r_{10},r_{11},r_{12},r_{13}, \\
    & r_{12},r_{13},r_{14},r_{15},r_{16},r_{17}, \\
    & r_{16},r_{17},r_{18},r_{19},r_{20},r_{21},\\
    & r_{20},r_{21},r_{22},r_{23},r_{24},r_{25},\\
    & r_{24},r_{25},r_{26},r_{27},r_{28},r_{29}, \\
    & r_{28},r_{29},r_{30},r_{31},r_{32},r_{1} \\
    \}
\end{split}
\end{equation}

Given the constraint from \autoref{eqn:a_from_e_r}, we thus present the desired trio of inputs for reverse engineering all bits of round key $K_1$ as Equations~\ref{eqn:a-1}-\ref{eqn:a-3}.
\begin{equation}\label{eqn:a-1}
\begin{split}
    a^1_{1..48} &= (000000\,000000\,000000\,000000\,000000\,000000\,000000\, 0000000)_2
\end{split}
\end{equation}
\begin{equation}\label{eqn:a-2}
\begin{split}
    a^2_{1..48} &= (001000\,001000\,001000\,001000\,001000\,001000\,001000\, 001000)_2
\end{split}
\end{equation}
\begin{equation}\label{eqn:a-3}
\begin{split}
    a^3_{1..48} &= (100010\,100010\,101000\,000101\,010001\,010001\,010101\,010010)_2
\end{split}
\end{equation}

However, we cannot just present the desired inputs at location $a$. 
One option is to \emph{scan in} the value to the $R$ register using the scan chain and \autoref{eqn:a_from_e_r} to determine the $R$ bits. However, depending on the circuit, it may have additional barriers to scanning in values as opposed to scanning them out.
Another option is to determine the value from the perspective of the initial input. Using the values within the Initial Permutation table, we can derive equations for $L_0$ and $R_0$, as presented in \autoref{eqn:des-l} and \autoref{eqn:des-r}.
\begin{equation}\label{eqn:des-l}
\begin{split}
L_0 = l_{1..32}=\{~~~~& \\
    & i_{58}, i_{50}, i_{42}, i_{34}, i_{26}, i_{18}, i_{10}, i_{2}, \\
    & i_{60}, i_{52}, i_{44}, i_{36}, i_{28}, i_{20}, i_{12}, i_{4}, \\
    & i_{62}, i_{54}, i_{46}, i_{38}, i_{30}, i_{22}, i_{14}, i_{6}, \\
    & i_{64}, i_{56}, i_{48}, i_{40}, i_{32}, i_{24}, i_{16}, i_{8}, \\
    \}
\end{split}
\end{equation}
\begin{equation}\label{eqn:des-r}
\begin{split}
R_0 = r_{1..32}=\{~~~~& \\
    & i_{57}, i_{49}, i_{41}, i_{33}, i_{25}, i_{17}, i_{9}, i_{1}, \\
    & i_{59}, i_{51}, i_{43}, i_{35}, i_{27}, i_{19}, i_{11}, i_{3}, \\
    & i_{61}, i_{53}, i_{45}, i_{37}, i_{29}, i_{21}, i_{13}, i_{5}, \\
    & i_{63}, i_{55}, i_{47}, i_{39}, i_{31}, i_{23}, i_{15}, i_{7} \\
    \}
\end{split}
\end{equation}

Given this, we can substitute \autoref{eqn:des-r} into \autoref{eqn:a_from_e_r} to produce an equation for $a$ based only upon original input $i$, \autoref{eqn:des-a-from-r}:

\begin{equation}\label{eqn:des-a-from-r}
\begin{split}
a_{1..48}=\{~~& \\
    & i_{7},i_{57},i_{49},i_{41},i_{33},i_{25}, \\
    & i_{33},i_{25},i_{17},i_{9},i_{1},i_{59}, \\
    & i_{1},i_{59},i_{51},i_{43},i_{35},i_{27}, \\
    & i_{35},i_{27},i_{19},i_{11},i_{3},i_{61}, \\
    & i_{3},i_{61},i_{53},i_{45},i_{37},i_{29},\\
    & i_{37},i_{29},i_{21},i_{13},i_{5},i_{63},\\
    & i_{5},i_{63},i_{55},i_{47},i_{39},i_{31}, \\
    & i_{39},i_{31},i_{23},i_{15},i_{7},i_{57} \\
    \}
\end{split}
\end{equation}

Thus, from \autoref{eqn:des-l} and \autoref{eqn:des-a-from-r} we can produce the desired input $i$ for each of our inputs $a^{1..3}$ (Equations~\ref{eqn:a-1}-\ref{eqn:a-3}).
These are presented in \autoref{eqn:special-inputs}.
\begin{equation}\label{eqn:special-inputs}
\begin{split}
i^1_{1..64} &=(0000000000000000)_{16} \\
i^2_{1..64} &=(0000AA000000AA00)_{16} \\
i^3_{1..64} &=(8220000A8002200A)_{16}
\end{split}
\end{equation}

We can verify this answer in Python by presenting these inputs and then computing the value at point $a$ and checking that it matches the desired $a$-values from Equations~\ref{eqn:a-1}-\ref{eqn:a-3}.

\begin{lstlisting}[style=PythonStyle,caption={Code to prepare the trio of special inputs and determine the $a$ values for those inputs},label={lst:special-inputs}]
#Use three specially crafted inputs to determine the unique round key R1
# they ensure L1 is 0, and R1 has a special value in it
special_inputs = ["0000000000000000",
                  "0000AA000000AA00",
                  "8220000A8002200A"]

#Permute the special_inputs to compute the values that will be XORed with the
# round-key R1 before the s-boxes (i.e. compute the value 'a' in Fig. 2)
special_inputs_at_pt_a = []
for i in range(len(special_inputs)):
    after_ip = des.permute(des.hex2bin(special_inputs[i]), des.INITIAL_PERM, 64)
    l0 = after_ip[:32]
    r0 = after_ip[32:]
    r0_expanded = des.permute(r0, des.EXPANSION_FUNC, 48)
    r0_expanded_list = []
    for i in range(48):
        r0_expanded_list.append(int(r0_expanded[i],2))
    special_inputs_at_pt_a.append(r0_expanded_list)

#Example output
#For each value in special_inputs_at_pt_a,
# print it in blocks of 6 bits
for i in range(len(special_inputs_at_pt_a)):
    print("Special input %i (%s): " % (i, special_inputs[i]))
    print("Value at pt. a: ", end="")
    for j in range(len(special_inputs_at_pt_a[i])):
        if(j % 6 == 0):
            print(" ", end="")
        print("%d" % special_inputs_at_pt_a[i][j], end="")
    print()
exit()

\end{lstlisting}

\begin{lstlisting}[style=TextStyle,caption={Example Output for Listing~\ref{lst:special-inputs}},label={lst:special-inputs-output}]
Special input 0 (0000000000000000): 
Value at pt. a:  000000 000000 000000 000000 000000 000000 000000 000000
Special input 1 (0000AA000000AA00): 
Value at pt. a:  001000 001000 001000 001000 001000 001000 001000 001000
Special input 2 (8220000A8002200A): 
Value at pt. a:  100010 100010 101000 000101 010001 010001 010101 010010
\end{lstlisting}

In practice, we cannot observe the values at point $c$ (i.e. immediately after the scan-chains) directly.
Instead, we must scan out value $e$ from the $R$ register after the first round is completed (i.e. $R$ is $R_1$) and perform the inverse operations to return it to value $c$.
Firstly, we must undo the exclusive-or operation against $L_0$ to compute $d$ (\autoref{eqn:d}).
To simplify this, we set $L_0$ to be all zeros using \autoref{eqn:des-l} when setting the input $i$. This makes $d = e$.
We then need to perform the inverse of permutation $P$.
This is a straight-through permutation (i.e. every bit in input appears in different location in output), and so inverting it is straightforward. We present this as \autoref{eqn:des-d}.

\begin{equation}\label{eqn:des-d}
\begin{split}
d_{1..32}=\{~~& \\
    & c_{16}, c_{7}, c_{20}, c_{21}, \\
    & c_{29}, c_{12}, c_{28}, c_{17}, \\
    & c_{1}, c_{15}, c_{23}, c_{26}, \\
    & c_{5}, c_{18}, c_{31}, c_{10}, \\
    & c_{2}, c_{8}, c_{24}, c_{14}, \\
    & c_{32}, c_{27}, c_{3}, c_{9}, \\ 
    & c_{19}, c_{13}, c_{30}, c_{6}, \\
    & c_{22}, c_{11}, c_{4}, c_{25} \\
    \}
\end{split}
\end{equation}

This is presented in code in \autoref{lst:special-inputs-pt-c}.

\begin{lstlisting}[style=PythonStyle,caption={Code to compute the value at point $c$ for each of the special inputs},label={lst:special-inputs-pt-c}]
special_results_after_sbox_pt_c = []
#For each of the 3 special inputs
for special_input in special_inputs:
    #Run 3 rounds of the encryption over the special input (i.e. determine L1, R1)
    (_, scans) = dut.RunEncryptOrDecrypt(special_input, True, 3)

    #Using the scan chain layout we computed earlier, extract the values of L1 and R1 registers
    (l_reg, r_reg) = read_scan_l_r(left_r_scan_indices, right_r_scan_indices, scans[2])
   
    #Undo the P permutation to get the values directly emitted from the SBox 
    # (i.e. the values at point 'c' in Fig. 2)
    special_result = [None]*32
    for i in range(32):
        special_result[des.P_PERM[i]-1] = r_reg[i]
    special_results_after_sbox_pt_c.append(special_result)

#(For testing only)
#For each value in special_results_after_sbox_pt_c,
# print it in blocks of 4 bits
for i in range(len(special_results_after_sbox_pt_c)):
    print("Special input %i (%s): " % (i, special_inputs[i]))
    print("Value at pt. c: ", end="")
    for j in range(len(special_results_after_sbox_pt_c[i])):
        if(j % 4 == 0):
            print(" ", end="")
        print("%d" % int(special_results_after_sbox_pt_c[i][j]), end="")
    print()
exit()

\end{lstlisting}

\begin{lstlisting}[style=TextStyle,caption={Example Output for Listing~\ref{lst:special-inputs-pt-c}},label={lst:special-inputs-pt-c-output}]
Special input 0 (0000000000000000): 
Value at pt. c:  1111 1000 1110 0011 0001 1011 1010 0000
Special input 1 (0000AA000000AA00): 
Value at pt. c:  0011 0011 0101 1010 0110 0100 1100 1100
Special input 2 (8220000A8002200A): 
Value at pt. c:  1010 1001 0000 1001 1010 0101 1010 0011
\end{lstlisting}

We can now use the values at point $c$ along with the values at point $a$ to determine the possible s-box values according to the code in \autoref{lst:reverse-sbox}. This is presented in \autoref{lst:possible-s-box-round-keys}.

\begin{lstlisting}[style=PythonStyle,caption={Code to determine possible round-key values for each s-box given each of the three special inputs},label={lst:possible-s-box-round-keys}]
#For each of the s-box special results at point c, 
# use the function sboxes_output_to_possible_inputs() 
# to determine the possible key inputs given the input
# to the xor at point 'a' in Fig. 2.
sbox_possible_key_values = []
for i in range(len(special_results_after_sbox_pt_c)):
    sbox_possible_key_values.append(
        sboxes_output_to_possible_inputs(
            special_results_after_sbox_pt_c[i], special_inputs_at_pt_a[i]
        )
    )

#Example output
#Print the possible key values for each of the s-boxes for each of the 3 special inputs
for i in range(len(sbox_possible_key_values)):
    print("Special input %i (%s): " % (i+1, special_inputs[i]))
    print("Possible key values...")
    for j in range(len(sbox_possible_key_values[i])):
        print(" for s-box %i: " % (j+1), end="")
        for k in range(len(sbox_possible_key_values[i][j])):
            for l in range(len(sbox_possible_key_values[i][j][k])):
                print("%d" % sbox_possible_key_values[i][j][k][l], end="")
            print(' ', end="")
        print()
exit()
\end{lstlisting}

\begin{lstlisting}[style=TextStyle,caption={Example Output for Listing~\ref{lst:possible-s-box-round-keys}},label={lst:possible-s-box-round-keys-output}]
Example output:
Special input 1 (0000000000000000): 
Possible key values...
 for s-box 1: 001010 000011 110000 100001 
 for s-box 2: 000100 001101 110010 100011 
 for s-box 3: 000110 010111 111100 110101 
 for s-box 4: 000110 001111 110100 100001 
 for s-box 5: 000110 001111 100100 101001 
 for s-box 6: 011110 011011 111100 110001 
 for s-box 7: 011010 001111 110000 101101 
 for s-box 8: 011010 011001 110000 110111 
Special input 2 (0000AA000000AA00): 
Possible key values...
 for s-box 1: 011000 010101 110000 111101 
 for s-box 2: 000100 001001 110010 100001 
 for s-box 3: 000110 011101 110000 110011 
 for s-box 4: 000110 010011 101000 100001 
 for s-box 5: 000110 010111 110000 111001 
 for s-box 6: 011110 001101 111100 101001 
 for s-box 7: 011010 011111 100000 110111 
 for s-box 8: 010100 011001 100010 111011 
Special input 3 (8220000A8002200A): 
Possible key values...
 for s-box 1: 110000 110011 011000 011011 
 for s-box 2: 110010 111001 011010 011101 
 for s-box 3: 101010 101101 000110 001111 
 for s-box 4: 001001 011010 100001 110100 
 for s-box 5: 011011 000110 111001 101000 
 for s-box 6: 001101 011110 110111 111010 
 for s-box 7: 001111 011010 100101 111000 
 for s-box 8: 000110 011001 101000 101011 
\end{lstlisting}

For each s-box, there is only one possible key which will be present in the outputs for each of the three special inputs.
For example, for s-box $S_1$, the only possible key value which is present for all inputs 1-3 is value $(110000)_2$ (see Lines 31, 41, 51).
That means that the round key $K_1$ must begin with these six bits.
We can determine each of these using the code in \autoref{lst:sbox_key_reduction}.

\begin{lstlisting}[style=PythonStyle,caption={Code to reduce number of possible key inputs by removing non-duplicates},label={lst:sbox_key_reduction}]
#Each of the sbox_possible_key_values is a list of lists of possible 
# key inputs for that sbox.
# Starting from the first input, remove any possibility that is not present 
# in the other inputs.
# (i.e. find the only input that is in all three sets of sbox possibilities)

possible_roundkey_bits_after_expansion = []
for sbox_index in range(8):
    possible_values = sbox_possible_key_values[0][sbox_index]
    for i in range(1,len(sbox_possible_key_values),1): #start at 1
        other_values = sbox_possible_key_values[i][sbox_index]

        #remove any elements from possible_values that is not present in other_values
        possible_values = [x for x in possible_values if x in other_values]        

    possible_roundkey_bits_after_expansion.append(possible_values)

#Example output
#Print the possible key values for each of the s-boxes after this removal step
# There should only be 1 possible set of bits per section of the key
print("Possible roundkeys")
for i in range(len(possible_roundkey_bits_after_expansion)):
    print("Bits %i-%i have %i possible value: " %
    for j in range(len(possible_roundkey_bits_after_expansion[i])):
        for k in range(len(possible_roundkey_bits_after_expansion[i][j])):
            print("%d" % possible_roundkey_bits_after_expansion[i][j][k], end="")
        print(" ", end="")
    print()
exit()
\end{lstlisting}

\begin{lstlisting}[style=TextStyle,caption={Example Output for Listing~\ref{lst:sbox_key_reduction}},label={lst:sbox_key_reduction-output}]
Possible roundkeys
Bits 1-8 have 1 possible value: 110000 
Bits 9-16 have 1 possible value: 110010 
Bits 17-24 have 1 possible value: 000110 
Bits 25-32 have 1 possible value: 100001 
Bits 33-40 have 1 possible value: 000110 
Bits 41-48 have 1 possible value: 011110 
Bits 49-56 have 1 possible value: 011010 
Bits 57-64 have 1 possible value: 011001 
\end{lstlisting}

At this point, there should be only one possible round key. However, to enable experimental modifications to these `special values', we will proceed as if we have not identified the round-key uniquely.
If there are more than one possible value for a given section of the round-key, we would need to take the cartesian product of the possible key bits in order to produce a set of possible keys.
This is presented in \autoref{lst:list-roundkey-1}. As noted in \autoref{lst:list-roundkey-1-output}, the round key is uniquely determined.

\begin{lstlisting}[style=PythonStyle,caption={List possible round keys $K_1$ (should only be 1 key).},label={lst:list-roundkey-1}]
#Convert the set of possible bits per round-key section
# to a set of possible round keys for the round1 key 
# by taking the cartesian product of the possibilities
# Note: there should only be one possible sbox input per sbox at this point,
# so the cartesian product should only return one element.
possible_roundkeys_round1 = []
for components in itertools.product(*possible_roundkey_bits_after_expansion):
    possible_roundkey_round1 = []
    for component in components:
        possible_roundkey_round1.extend(component)
    possible_roundkeys_round1.append(possible_roundkey_round1)

#Example output
#Print the possible roundkeys for round1
for possible_roundkey_round1 in possible_roundkeys_round1:
    possible_roundkey_val = 0
    for i in range(48):
        possible_roundkey_val |= (possible_roundkey_round1[i] << (47-i))
    print("Possible roundkey 1: %012X" % possible_roundkey_val)
exit()
\end{lstlisting}

\begin{lstlisting}[style=TextStyle,caption={Example Output for Listing~\ref{lst:list-roundkey-1}},label={lst:list-roundkey-1-output}]
Possible roundkey 1: C321A119E699
\end{lstlisting}

\emphbox{\textbf{Insight:} 
The three special inputs were well-designed to reverse the s-boxes uniquely and produce only one possible round key $K_1$. 
However, what happens if the special inputs are not well-designed, or if we used only one or two special inputs?
Consider experimenting with this.
}

\subsection{Attack Phase 3: Determine the original key}

Each round key contains 48-bits of the original 56-bit key (8 of the 64-bits in the 64-bit key are parity bits not used for encryption purposes). 
After determining round key $K_1$, we now have 48 bits of the key (positions derivable by undoing the PC2, shifts, and PC1 permutations).
The attack can now proceed in two different ways.
The first method is to perform a similar attack on $R_2$ and $R_3$ as depicted above, by using the scan chain to \emph{scan in} replacement values to the $L$ and $R$ registers after each round of encryption is completed.
However, this requires additional test infrastructure.

The second method notes that there are only 8-bits remaining of the key, giving only $2^8=256$ different key bit possibilities.
As such, we should be able to brute-force the values relatively easily.

In this section we will proceed with the second method, and brute force the remaining bits of the key.
We present this in code in \autoref{lst:listing-possible-keys}.
Note the possible-key list structure printed by Line 40 (example output on Line 70).
This has \texttt{[0]} or \texttt{[1]} in the position of `known' bits, \texttt{[0, 1]} in position of `unknown' bits, and \texttt{[None]} in position of the parity bits (every 8th bit).
As there are 8 `unknown' bits, there should be 256 possible keys generated from the cartesian product, which is confirmed for the example on Line 71.

\begin{lstlisting}[style=PythonStyle,caption={Determining the list of possible keys that could generate round key $K_1$.},label={lst:listing-possible-keys}]
possible_keys = []
#For each possible round1 key, derive every possible key that could have generated it
for possible_roundkey_round1 in possible_roundkeys_round1:
    #First, undo the PC2 permutation
    key1 = [None]*56
    for i in range(48):
        key1[des.KEY_PC2[i]-1] = possible_roundkey_round1[i]
    
    #Now undo the two half-key rotations by
    # right rotating each half of the key
    key1_left = key1[:28]
    key1_right = key1[28:]
    key1_left = key1_left[-1:] + key1_left[:-1]
    key1_right = key1_right[-1:] + key1_right[:-1]
    key1 = key1_left + key1_right

    #Now undo the PC1 permutation
    key = [None]*64
    for i in range(56):
        key[des.KEY_PC1[i]-1] = key1[i]

    #The format of the key is such that it has 64 bits, 
    # but only 48 of them are currently filled with values (the others are 'None')
    # We will create all possible keys by taking the cartesian product
    # of all possible values for the 8 unfilled key bits (ignoring parity bits).
    
    #Prepare the cartesian product by creating a list of lists of 
    # known and possible values for the unfilled key bits.
    combined_key_possibilities = []
    for i in range(64):
        if key[i] == None:
            if (i+1) % 8 != 0:
                combined_key_possibilities.append([0,1]) #Unknown key bit
            else:
                combined_key_possibilities.append([None]) #Parity bit
        else:
            combined_key_possibilities.append([key[i]]) #Known key bit
    
    print("Key possibilities that would generate round key 1:")
    print(combined_key_possibilities)
    
    for components in itertools.product(*combined_key_possibilities):
        #Combine all key bits into a single key
        possible_key_bits = []
        for component in components:
            possible_key_bits.append(component)

        #Calculate the parity bits
        for i in range(8):
            val_bits = possible_key_bits[i*8:(i+1)*8-1]
            parity_bit = 0
            for bit in val_bits:
                parity_bit ^= bit
            possible_key_bits[i*8+7] = parity_bit
        
        #Convert the binary list into a hex string
        key_val = 0
        for i in range(64):
            key_val |= (possible_key_bits[i] << (63-i))
        possible_key_val = '%016X' % key_val

        #Store the possible key
        possible_keys.append(possible_key_val)

#Example output
print("There are %d possible keys." % len(possible_keys))
\end{lstlisting}

\begin{lstlisting}[style=TextStyle,caption={Example Output for Listing~\ref{lst:listing-possible-keys}},label={lst:listing-possible-keys-output}]
Key possibilities that would generate round key 1:
[[0], [0], [0], [0], [1], [0, 1], [0, 1], [None], [0], [1], [0, 1], [0, 1], [1], [1], [1], [None], [0], [0], [1], [0], [1], [0], [1], [None], [1], [0], [0], [0], [0], [1], [1], [None], [1], [0], [0], [0], [1], [1], [0], [None], [1], [0], [0, 1], [1], [0], [0, 1], [0], [None], [0], [0, 1], [1], [0, 1], [1], [1], [0], [None], [1], [0], [1], [0], [0], [0], [0], [None]]
There are 256 possible keys.
\end{lstlisting}

The remaining step is to now perform the brute-force operation to check all the possible keys until the correct key is found. 
This is straightforward by creating new DES instances with specified keys (refer to \autoref{fig:python-des-diagram}). We will compare the ciphertexts that we generate to that generated in \autoref{lst:des-dut-instantiate}.
The brute-forcing code is presented in \autoref{lst:key-brute-force}.

\begin{lstlisting}[style=PythonStyle,caption={Brute forcing the key from the list of possibilities},label={lst:key-brute-force}]
print("Brute-force checking %d possible keys." % len(possible_keys))
for possible_key in possible_keys:
    pos_des = des.DESWithScanChain(force_key=possible_key)
    (test_ciphertext, _) = pos_des.RunEncryptOrDecrypt(test_code)
    if(test_ciphertext == check_ciphertext):
        print("Found the key. It is %s" % possible_key)
        break

print("Checking the answer. The embedded secret key was " + dut.key_hex)
if(possible_key == dut.key_hex):
    print("The two keys match, the attack is successful.")
\end{lstlisting}

\begin{lstlisting}[style=TextStyle,caption={Example Output for Listing~\ref{lst:key-brute-force}},label={lst:key-brute-force-output}]
Example output:
Brute-force checking 256 possible keys.
Found the key. It is 096F2B878D906CA0
Checking the answer. The embedded secret key was 096F2B878D906CA0
The two keys match, the attack is successful.
\end{lstlisting}

The attack successfully recovered the key used to generate the round keys in the hardware.
\subsection{Discussion}

\subsubsection{Design exploration}
\label{sec:design-explore}
In this case study we explored how to perform a scan chain attack on (emulated) hardware.
We presented relatively simple iterative hardware, but this attack would also work on more complex implementations.
For instance, in a fully pipelined architecture, each of the 16 DES rounds would be instantiated separately, meaning there would be 17 pairs of $L$ and $R$ registers (from $\{L_0, R_0\}$ to $\{L_{16}, R_{16}\}$).
While this would significantly increase the size of the scan chain, identifying the correspondence between each bit and its position in the scan chain is still possible, especially for the low registers~\cite{yang_scan_2004}.
$L_0$ and $R_0$ can be located first, by using the same methodology as presented earlier.
As $L_1=R_0$, identifying $L_1$ is also straightforward.
Given $R_1=L_0\oplus f(R_0,K_1)$ we can determine $R_1$ as follows.
We set $R_0$ to be a constant value; $K_1$ is already constant.
We then iterate through $L_0$, setting each bit to be $1$ in turn.
As $f(R_0,K_1)$ is constant, there will only be a 1-bit difference each time we stream out $R_0$, and that difference will be based on the position of the $1$ in $L_0$.
Now we have the positions of $L_0$, $R_0$, $L_1$, and $R_1$ and the key extraction can proceed as outlined in the previous subsection (using the brute force method). 

Our emulated hardware assumed that it took one clock cycle to load values into the input register.
For some DES implementations this might take a different number of cycles, or no cycles at all if there is no input register present.
However, using the scan chain it also possible to determine this.
Firstly, the length of the scan chain reveals how many register bits are present.
Secondly, by observing the scan chain after each clock tick, it is possible to determine when encryption has commenced.
This is because cryptographic algorithms such as DES display an `avalanche' effect~\cite{yang_scan_2004}, where small differences (i.e. 1-bit) will translate into larger changes in the subsequent rounds.
By observing for this, it is possible to determine when the encryption has started~\cite{yang_scan_2004}.

Our emulated hardware simplified some aspects of the reverse engineering by excluding control registers from the scan chain. 
If these were present, there would be additional and varying data bits in each cycle.
However, these are control-driven rather than data-driven.
As such, we can identify them by loading multiple different inputs to the hardware module and running the complete encryption/decryption cycle.
Any flip-flops in the control unit would have the same pattern for each input trace, no matter the input, and could thus be identified and discarded.
Further, we assume that a reset actually sets all internal register bits to zero.
This may not be the case, as some circuits have implementations that reset to unknown or random garbage values.
If this is the case, then instead of resetting the circuit using the random reset, the attacker should `reset' the circuit by loading a constant input and running the complete encryption procedure.
This would set all internal registers to a constant value, which can be the new fixed state of the chip rather than all bits being zero.

\subsubsection{Potential pedagogical exercises}
This case study presents a walk-through of an attack on a DES implementation, providing a sound basis for academic exercises. Consider modifying the design (e.g. using the suggestions in Section~\ref{sec:design-explore}) prior to attempting the attack. Also consider attacking a different encryption algorithm that is vulnerable to scan attacks, such as AES~\cite{ali_test-mode-only_2014,yang_secure_2006}.

\subsection{What's next in Scan Attacks?}

Given the ongoing need for correctness in IC manufacturing alongside the pressing demands for new technologies, processes, and increased complexity, the presence of scan chains in integrated circuits is a relative certainty. 
Preventing abuse of the scan chain is thus paramount.
Two broad categories of approaches are proposed~\cite{azar_cryptography_2021}: (1) preventing unauthorized access to the scan chains, (2) obfuscating the scan chain I/O. These are discussed further here.

\subsubsection{Preventing Unauthorized Access}

Blocking access to the scan chain seems an intuitive method for protecting it, for instance using blocking logic / gates. 
However, blocks may be able to be overcome, especially if they exist only at the physical layer. 
Visual analysis (using reverse engineering techniques) will be able to identify the scan chain in the embedded IC die.
Digital locking logic, meanwhile, may also be able to be bypassed: for instance, the technique R-DFS~\cite{guin_robust_2018} which controls the scan chain via so-called secure cells is defeated via an attack termed `shift-and-leak'~\cite{limaye_is_2019}.
Variants of this have all also shown weaknesses. 

An alternative method for protecting unauthorized viewing of secret data, introduced in~\cite{yang_secure_2006}, uses an approach termed `mirror key registers' (MKR). Here, the general approach is to prevent sensitive data (e.g. embedded keys) from entering the scan chain during the test mode (scanning). In the presented case study, this could be realized by using two sets of embedded keys in the DES hardware: one only for use during scan chain tests, and one for use during normal operation.

\subsubsection{Obfuscating the Scan Chain I/O}

A number of techniques have been proposed (and broken) for the protection of scan chain logic. 
An early attempt, termed a `flipped scan'~\cite{sengar_secured_2007}, statically inserts a number of inverters (NOT gates) between randomly selected scan chain flip-flops. However, this can be defeated by carefully analyzing the pattern of 1's and 0's after the reset condition~\cite{agrawal_scan_2008}.
Replacing the NOT gates with XOR gates and introducing feedback loops is another suggestion which has been broken~\cite{banik_improved_2013,banik_cryptanalysis_2014}.
As such, adding multiplexers with each XOR gate, to enable dynamic control of the scan chain, is also proposed: here, keys (known by the tester)~\cite{atobe_dynamically_2012} or randomness governed by a physically unclonable function~\cite{banik_cryptanalysis_2014} or linear feedback shift registers~\cite{zhang_dynamically_2017} are required for the scan chain to enter the correct configuration. However, these may still be broken using approaches such as ScanSAT~\cite{alrahis_scansat_2019}, which has also shown that it can break scan chain compression techniques.
Keys may be also used to fully scramble the order of the scan chain~\cite{hely_scan_2004}, although this introduces a relatively large layout overhead in the IC. 
However, any system which just reorders the bits of the scan chain will be crackable eventually~\cite{azar_cryptography_2021,cui_why_2017}.

\subsubsection{Future work on protecting scan chains}

Additional hardware used to protect scan chains may be a path forwards, for instance using a secondary encryption system to encrypt data before it is emitted from the hardware or during the scanning process. One such technique, using so-called `Encrypt Flip Flops', has shown promise in this area~\cite{karmakar_scan_2020,karmakar_encrypt_2018,paul_obviating_2022}, but may require the tester to have access to the encryption key, thus rendering the approach vulnerable to key losses. In addition, ScanSAT attacks have also shown they may break these approaches~\cite{alrahis_scansat_2021}.
Future work will no doubt continue this cat-and-mouse game exploring new protections for scan chains, as well as determine the weaknesses of those protections. %

\clrpg{}

\section{Case Study: Digital Hardware Intellectual Property Protection\label{sec:locking}}

\subsection{Overview}
In this section, we will work towards building an intuition about digital hardware intellectual property (IP) protection, specifically through logic locking,\footnote{This is also seen in literature as logic obfuscation, logic encryption, redaction, and so on.} a family of techniques that continues to evolve since first appearing in 2008~\cite{roy_epic_2008}. 

\subsubsection{Setting}
As one can intuit, hardware IP is extremely valuable. 
Creators of IP (or IP vendors) want to manage access of the IP within the context of licensing agreements and to avoid threats such as reverse-engineering, where malicious parties want to understand an IP to the level where they might be able to illicitly modify the design (say, to insert a Hardware Trojan) or piracy (where designs are stolen, cloned, and passed off as genuine or as a competing product). 
Ultimately, the designer (acting as a \textbf{defender} of the IP) wants to ensure that only legitimate users (or licensees) can use the IP as intended while other entities cannot. 

\subsubsection{Threat Models \label{sec:lock-tut}}
In general terms, we can consider the following entities (and their employees) as being involved in the chip design flow: the \textbf{IP vendor}, the \textbf{system-on-chip integrator}, the \textbf{foundry}, the \textbf{test facility}, and the \textbf{end user}. 
Usually, we want to protect IP from potential adversaries: untrusted foundries, test facilities, or compromised/illegitimate end users, assuming that they have been compromised in some way (see survey papers like~\cite{chakraborty_keynote_2020,beerel_towards_2022} for more discussion about the threat model). 
Their aim is to do whatever they can to recover the IP; usually defined as gaining enough knowledge of the design so that they can get the IP to produce functionally correct outputs. 
We will see more precisely what this means in the context of logic locking in the next section. 
We usually \textbf{assume that the adversary can recover the IP's gate-level netlist}. 
Without any protection, an adversary can work out what different parts of the gate-level netlist do, work out what parts to modify if they desire, and re-use the IP. 

Given this setting, there are two typical variations of the threat model. 
The \textbf{Oracle}-based model assumes that the adversary has in their possession an instance of the IP that produces \textbf{functionally correct outputs}---perhaps (legitimately) acquired on the open market. 
Within the \textbf{Oracle}-based approach, there are additional possible assumptions, including whether scan-chain access is available or if the Oracle's internal behavior can be observed (e.g., tamper-proof memory used or packaging-level protections to mitigate imaging). 
The \textbf{Oracle-less} model assumes that the adversary only has the gate-level netlist and is unable to gain information regarding correct functional input/output behavior. 

\emphbox{\textbf{Insight:} As with any security problem, it is up to each practitioner to decide for themselves what assumptions are reasonable in terms of threat modeling. Take a moment to consider what adversary capabilities you think are realistic. In this tutorial, we will focus our attention on the basic locking of \textbf{combinational logic circuits} and \textbf{Oracle}-based analysis. From this foundation you should be well-equipped to engage with the recent advances in logic locking research. }

\subsection{Logic Locking: A ``family'' of approaches}

Given the aforementioned threat model, \textbf{logic locking} seeks to transform a design in such a way that an adversary cannot simply copy or reverse-engineer a design\footnote{Recent work by Beerel et al.~\cite{beerel_towards_2022} attempts to formalize the aims of logic locking, but we will take the intuitive understanding as sufficient for the purposes of an introduction to the topic in this tutorial paper.}. 
In other words, we want to \textbf{change the design by inserting or replacing some of its logic} so that the IP is associated with some \textit{secret} that can be shared between the IP vendor and trusted legitimate end-user that will ``unlock'' the IP's true functionality. The secret is typically in the form of some sort of ``key'' input. 

\begin{figure}[h]
    \centering
    
    \subcaptionbox{A simple circuit\label{simplecirc}}[.4\textwidth]{\includegraphics[width=0.4\textwidth]{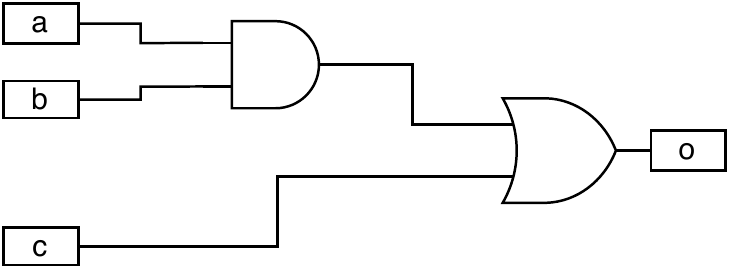}}%
    \hfill
    \subcaptionbox{A locked circuit\label{lockedcirc}}[.4\textwidth]{\includegraphics[width=0.4\textwidth]{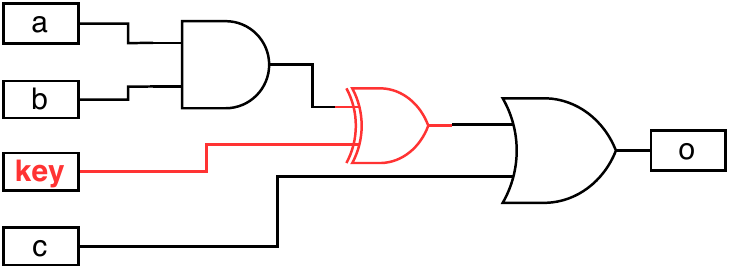}}
    \caption{An example of logic locking}
    \label{fig:original-netlist}
\end{figure}

\begin{table}[h]
\caption{Truth Table for the circuit in \autoref{fig:original-netlist}. Notice the differences in the output `o' when the key is set to 0 or 1. }
\label{tab:original-table}
\begin{tabular}{@{}|ccc|c||ccc|c||ccc|c|@{}}

\multicolumn{4}{c}{Original} & \multicolumn{4}{c}{key = 0} & \multicolumn{4}{c}{key = 1} \\ \hline
a     & b     & c     & o     & a     & b     & c     & o    & a          & b          & c          & o          \\ \hline
0     & 0     & 0     & 0     & 0     & 0     & 0     & 0    & \textbf{0} & \textbf{0} & \textbf{0} & \textbf{1} \\
0     & 0     & 1     & 1     & 0     & 0     & 1     & 1    & 0          & 0          & 1          & 1          \\
0     & 1     & 0     & 0     & 0     & 1     & 0     & 0    & \textbf{0} & \textbf{1} & \textbf{0} & \textbf{1} \\
0     & 1     & 1     & 1     & 0     & 1     & 1     & 1    & 0          & 1          & 1          & 1          \\
1     & 0     & 0     & 0     & 1     & 0     & 0     & 0    & \textbf{1} & \textbf{0} & \textbf{0} & \textbf{1} \\
1     & 0     & 1     & 1     & 1     & 0     & 1     & 1    & 1          & 0          & 1          & 1          \\
1     & 1     & 0     & 1     & 1     & 1     & 0     & 1    & \textbf{1} & \textbf{1} & \textbf{0} & \textbf{0} \\
1     & 1     & 1     & 1     & 1     & 1     & 1     & 1    & 1          & 1          & 1          & 1          \\ \hline
\end{tabular}%
\end{table}

For example, consider the circuit in \autoref{fig:original-netlist} and its associated truth table in \autoref{tab:original-table}. 
Imagine that this is a valuable IP and we want to transform the design (i.e., ``lock'' it) such that the adversary cannot get the benefit of correct input/output behavior. 
The simplest modification we could do the circuit is to insert an \texttt{XOR} gate somewhere in the design and connecting it to a ``key input'' as in \autoref{lockedcirc}. 
Consider the truth table in \autoref{tab:original-table} when \texttt{key} is set to $0$ and when it is set to $1$---notice that some of the rows are different to the original when $key=1$. 
If the correct value of the key input is kept secret, an adversary that makes an incorrect guess of the key input will end up with a circuit that does not behave correctly for all input combinations. 
This is the basic premise of the first logic locking approach proposed by Roy et al. in 2008~\cite{roy_epic_2008}: randomly add \texttt{XOR} and \texttt{XNOR} gates to various points of the design attached to new key inputs. 
\textit{As an exercise, you can create your own 3-input, 1-output circuit with 4-5 logic gates, construct its truth table, and then examine what happens when you ``lock'' the design with \texttt{XOR} and \texttt{XNOR} insertion. }

\emphbox{\textbf{Insight:} One way to think about the general premise of logic locking is to add ``additional'' or ``surplus'' functionality to the design, from which only a subset is actually useful. In our basic example, we added a new input and a gate, which actually expands the Boolean function from a 3-input to 4-input function. 
Added parts used in logic locking include new (key) inputs (and thus, new parts of the truth table) or new states and transitions (such as by adding memory elements like flip-flops). 
The useful (protected) part of the design should be hard to discover or use, unless, of course, you are authorized to by receiving the requisite secret material.}

Since 2008, there has been several back-and-forth exchanges in the academic community with proposed attacks and countermeasures~\cite{tan_benchmarking_2020,chakraborty_keynote_2020}. 
Check out the references for a selection of related work. 
For this tutorial, we will keep ourselves occupied with random logic locking (RLL) (basic \texttt{XOR}/\texttt{XNOR} locking). 

\subsection{Early analyses} 
Early analyses of logic locking has had a close link with VLSI testing ideas, so to start to get a flavor of the type of analyses that is possible, let us consider the idea of the \textbf{sensitization attack}~\cite{rajendran_security_2012}, where an adversary has access to an activated chip (i.e., a locked design which has had the secret material loaded to unlock the functionality).  

Consider again the simple design of \autoref{lockedcirc}. 
We know the structure of the design, but we do not know the key input's intended value---the idea here is to sensitize the path so the key input value in the activated chip can be observed directly on the output, which we can do by choosing appropriate test input vectors. 

\begin{figure}[h]
    \centering
    {\includegraphics[width=0.4\textwidth]{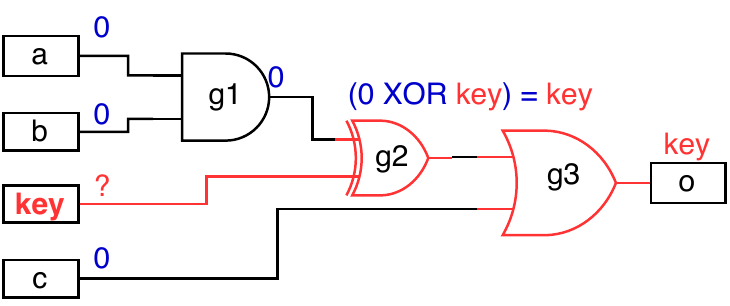}}
    \caption{Sensitization attack}
    \label{fig:sens}
\end{figure}

For example, as we can see in \autoref{fig:sens}, one can set the inputs ``abc'' to ``000'' so that the actual value of key can be seen on the output. By setting $c = 0$, we ensure the output of gate g3 reflects the output of gate g2. By setting either a or b to $0$, we can guarantee that the output of gate g1 is 0, thus making the output of gate g2 the value of the \texttt{key} input. 
Typical test generation algorithms can help adversaries find the test vectors needed to sensitize the design~\cite{rajendran_security_2012,duvalsaint_characterization_2019}. 

As one might expect, a defender that knows that this attack is possible can try to respond by adding multiple key inputs and gates such that it is not possible to sensitize an individual key input~\cite{rajendran_security_2012}, i.e., the choice of where to insert locking logic needs to be carefully considered. 
Recent work has shown that the sensitization and other attacks that use automatic test pattern generation (ATPG) have shown success on various logic locking flavors~\cite{duvalsaint_characterization_2019,duvalsaint_characterization_2019-1}. 
This kind of adversarial back-and-forth or ``cat-and-mouse'' perspective has driven a lot of progress in logic locking~\cite{beerel_towards_2022,tan_benchmarking_2020,chakraborty_keynote_2020}. 

Next, we will take a closer look at one of the most influential attacks in the logic locking literature: the ``SAT attack''~\cite{subramanyan_evaluating_2015}. 
Understanding the SAT attack provides a good foundation for understanding the motivations for and intuitions around many proposed logic locking approaches, such as cyclic locking~\cite{zhou_cycsat_2017}, stripped functionality logic locking~\cite{yasin_provably-secure_2017}, its descendants~\cite{sengupta_truly_2020}, and attacks (e.g.,~\cite{shamsi_kc2_2019,shamsi_icysat_2019,shen_besat_2019,shamsi_appsat_2017}). 

\subsection{The SAT attack}
In this section, we will do a step-by-step walk-through of the concepts and application of the attack by Subramanyan et al.~\cite{subramanyan_evaluating_2015}. 

\subsubsection{The building blocks}
The SAT attack builds on a few key ideas, including the Boolean satisfiability problem, the construction of miter circuits, and the identification of distinguishing input patterns. 
The first idea is the \textbf{Boolean satisfiability problem}, often referred to as \texttt{SAT}. 
Given a Boolean formula, \texttt{SAT} is the problem of determining if variables in the formula can be set to \texttt{TRUE} and \texttt{FALSE} such that the formula as a whole evaluates to \texttt{TRUE}. 
If so, the formula is \textit{satisfiable} (or SAT), otherwise, it is \textit{unsatisfiable} (or UNSAT). 
While \texttt{SAT} is an NP-complete problem, several heuristic algorithms exist that can be used to solve \texttt{SAT} instances~\cite{gu_algorithms_1996}. 
So, if we can produce a Boolean formula, we can give it to a SAT solver to try to find the variable assignments that make the formula evaluate to \texttt{TRUE}, if the formula is satisifiable. 
For example, the formula $a \land b$\footnote{The notation $\land$ means logical conjuction (\texttt{AND}). $\lor$ means logical disjunction (\texttt{OR}).} is satisfiable, as the formula evaluates to \texttt{TRUE} when $a = 1$ and $b = 1$. 

\begin{figure}[h]
    \centering
    
    \subcaptionbox{\label{fig:miter}}[.46\textwidth]{\raisebox{7mm}{\includegraphics[width=0.4\textwidth]{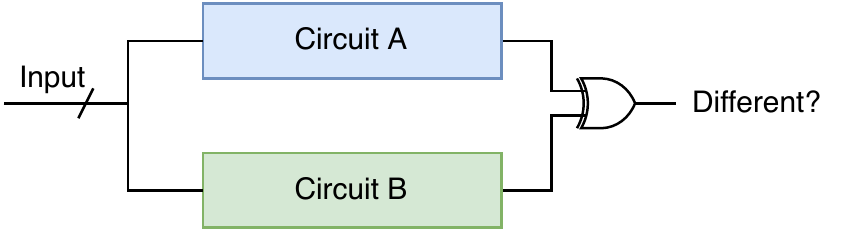}}}%
    \subcaptionbox{\label{fig:miter-attack}}[.46\textwidth]{\includegraphics[width=0.4\textwidth]{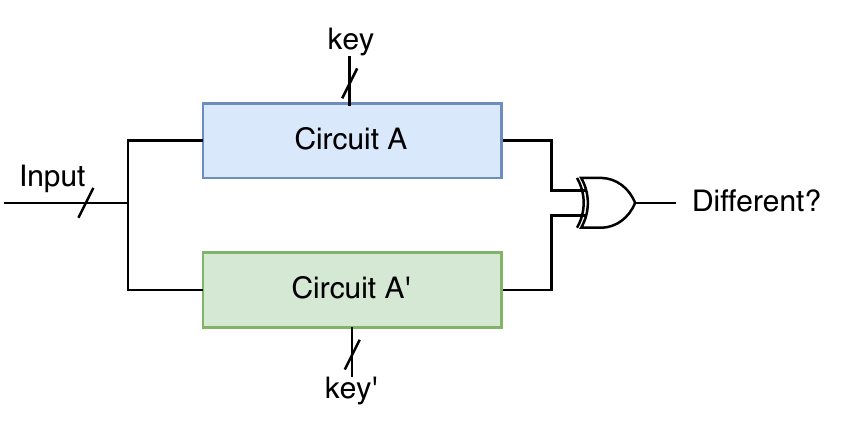}}
    \caption{(a) A general miter circuit and (b) the starting miter for the SAT attack. }
    \label{fig:miter-examples}
\end{figure}

The next thing we need to understand is the idea of a \textbf{miter circuit}. 
A miter circuit is a circuit comprising two circuits that are fed the same input, as shown in~\autoref{fig:miter}; we can check to see if the output of the circuits match for any given input. 
Intuitively, the two circuits are equivalent if they produce the same output as each for any input. 
The miter circuit is especially useful if we use a SAT solver because we can write a Boolean formula representing the miter circuit (including the output comparison) to feed into the SAT solver. 
The SAT solver then tries to answer the question: "Is there any input where the circuit outputs are different"? 
A SAT solver that returns SAT has found an input where the output of the two circuits disagrees. 

We can use the miter circuit to perform formal equivalence checking of two combinational circuits: say we have two circuits that we \textit{think} have the same behavior, we can connect them together as a miter (same inputs, \texttt{XOR}'d output) and use a SAT solver to see if the miter is satisfiable. 
The circuits are equivalent if the solver returns UNSAT; i.e., the solver cannot find any input that causes the two circuits to produce different outputs (i.e.,  \texttt{XOR} at the end will always be 0 (\texttt{FALSE})). 

This leads us to the third idea: the idea of a distinguishing input pattern. 
Let us now consider a miter circuit that is built with two copies of a logic locked design, as in \autoref{fig:miter-attack}. 
As before, the same input is fed into each circuit copy. 
When we feed this miter circuit into a SAT solver, we are essentially asking the same question as before: "\textit{Is there any input where the circuit outputs are different?}" -- except, in this case, we have separate \texttt{key} and \texttt{key'} inputs to each copy of the locked circuit. 
Given that the two circuits in the miter are exactly the same, the only way in which a SAT solver will be able to make the miter circuit formula satisfiable (if it is satisifiable) is to find an input and different values of \texttt{key} and \texttt{key'} that will make the different copies present different outputs. 
In other words, the solver will find a \textbf{distinguishing input pattern} (DIP) which means that the value (variable assignments) of \texttt{key} and \texttt{key'} that are found belong to two different classes of keys that make the locked circuit behave differently for that DIP. 

\subsubsection{The algorithm}
We can now put together the aforementioned ideas into an an attack on logic locked designs when the adversary has access to an Oracle. 
Let us assume that we created the miter circuit using two copies of the locked design and that we have fed this to a SAT solver that returns SAT. 
The solver has given as a DIP and two possible outputs for that DIP. 
\textit{Which output is correct?} 
With access to an Oracle, we can simply ask it which is the correct output for that DIP. 
Crucially, because we know which output is correct for that DIP, we can revise the miter circuit Boolean formula that we give to the SAT solver to include the new information -- i.e., that for the DIP, the output must be so. 
The formula can be fed into the SAT solver again, which will attempt to find another DIP. 
Each time we find a DIP, we can query the Oracle, modify the formula to take into account what the correct behavior should be for that DIP, and repeat; the solver prunes the key space iteratively. 
In each iteration, we effectively ask the solver to answer the question: \textit{"Is there any input where the circuit outputs are different, given that when the inputs are \{previous DIPs\} the corresponding outputs must be \{the correct outputs from the Oracle\}"}?
Eventually, the solver will return UNSAT when it can no longer find any DIPs; what is left will let us identify a key in the equivalence class of correct keys. 

Taken all together, \autoref{alg:sat_attack} presents Subramanyan et al.'s ``Logic Decryption Algorithm''~\cite{subramanyan_evaluating_2015}, where $C(I,K,O)$ is the Boolean input-key-output relation (i.e., logic locked circuit represented as a Boolean formula), $I$ is a vector of inputs, $K$ is a vector of key inputs, $O$ is a vector of outputs, and $oracle()$ is a function that represents querying the Oracle. Subscripts represent each iteration of the algorithm. 

\begin{algorithm}[h] \small
\SetAlgoLined
\SetKwData{Left}{left}\SetKwData{This}{this}\SetKwData{Up}{up}\SetKwFunction{Union}{Union}\SetKwFunction{FindCompress}{FindCompress}\SetKwInOut{Input}{Input}\SetKwInOut{Output}{Output}

\Input{Locked circuit C, Oracle}
\Output{The values of K}
 i = 1 \;
 $F_1 = C(I, K_1, O_1) \land C(I, K_2, O_2)$\;
 \While{$sat[F_i \land (O_1 \neq O_2)]$}{
    $I^d_i =$ a DIP value that satisfy $[F_i \land (O_1 \neq O_2)]$\;
    $O^d_i = oracle(I^d)$\;
    $F_{i+1} = F_i \land C(I^d_i, O^d_i, K_1) \land C(I^d_i, O^d_i, K_2)$\;
    $i = i + 1$\;
 }
 $K =$ the value of $K$ in the sat assignment of $F_i \land (O_1 \equiv O_2)$\; 
 \caption{SAT Attack Algorithm from~\cite{subramanyan_evaluating_2015} (variable names changed )}
 \label{alg:sat_attack}
\end{algorithm}

\subsubsection{Worked example} 
\begin{figure}[h]
    \centering
    \includegraphics[width=0.8\textwidth]{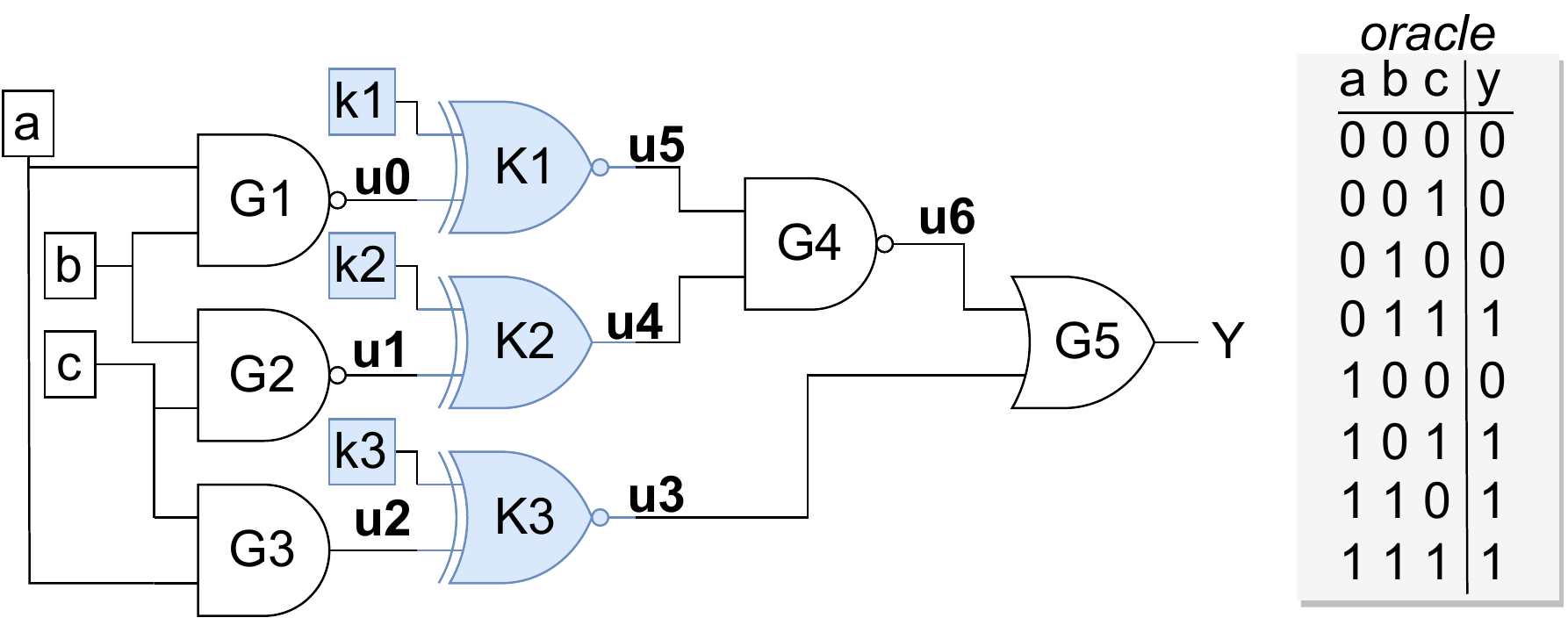}
    \caption{Logic locked example design with its "Oracle" as a truth table}
    \label{fig:lock-example}
\end{figure}

To see more clearly how the SAT attack algorithm process works, let us try a worked example on the locked design shown in \autoref{fig:lock-example}. 

\begin{figure}[h]
    \centering
    \includegraphics[width=0.8\textwidth]{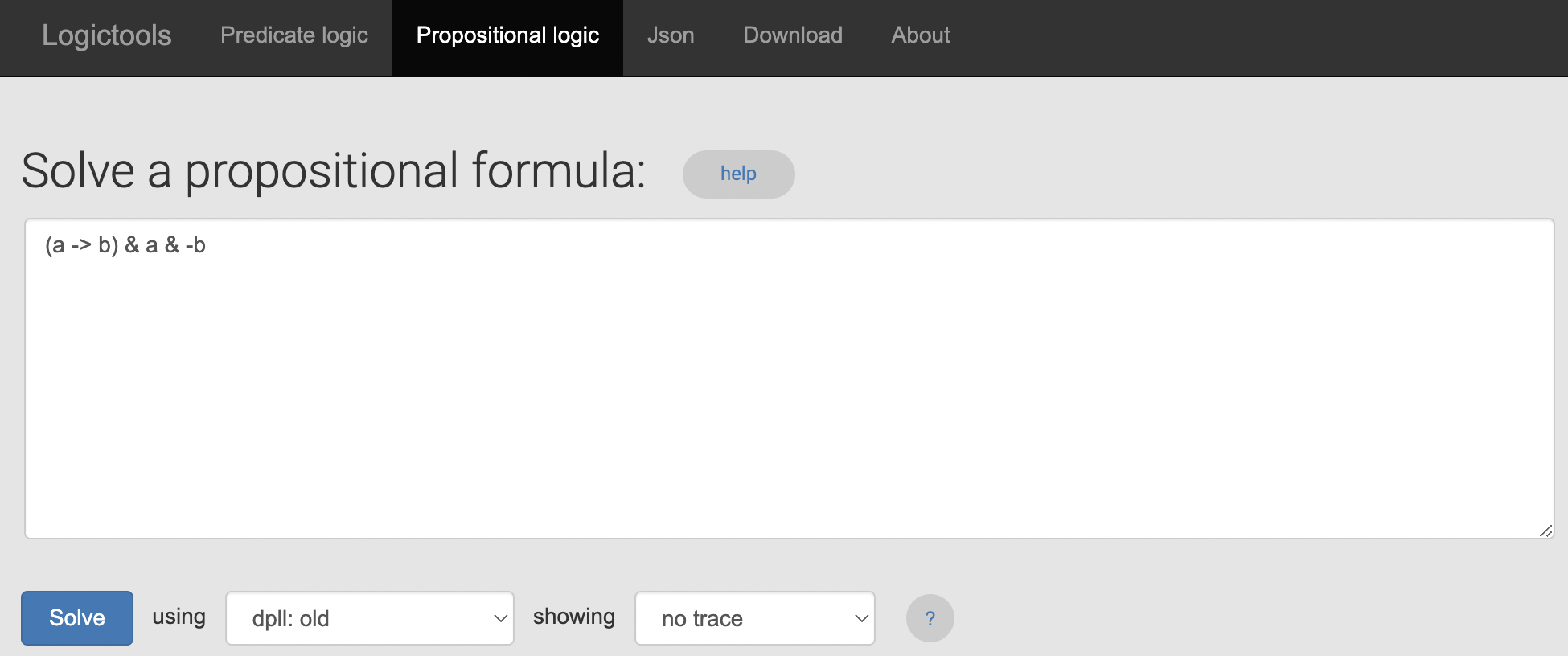}
    \caption{The web browser-based logictools SAT solver from~\cite{tammet_logictools_2022}}
    \label{fig:logictools}
\end{figure}

For this example, we will use a simple, Web browser-based SAT solver called logictools~\cite{tammet_logictools_2022} with its corresponding syntax for representing propositional formulas\footnote{\texttt{+} is \texttt{XOR}, \texttt{<=>} is equivalence, \texttt{\&} is \texttt{AND}, $\sim$ or \texttt{-} is negation.}. We will use the \texttt{dpll:old} engine in this walk-through, as shown in \autoref{fig:logictools}. 
Typically, SAT solver tools ingest formulas in conjunctive normal form (CNF) which can be produced from propositional formulas using the Tseytin transformation~\cite{siekmann_complexity_1983}; logictools can ingest propositional formulas directly (and performs the transformation internally), so we will stick with propositional formulas for clarity in this example.

\begin{figure}[h]
    \centering
    \includegraphics[width=0.9\textwidth]{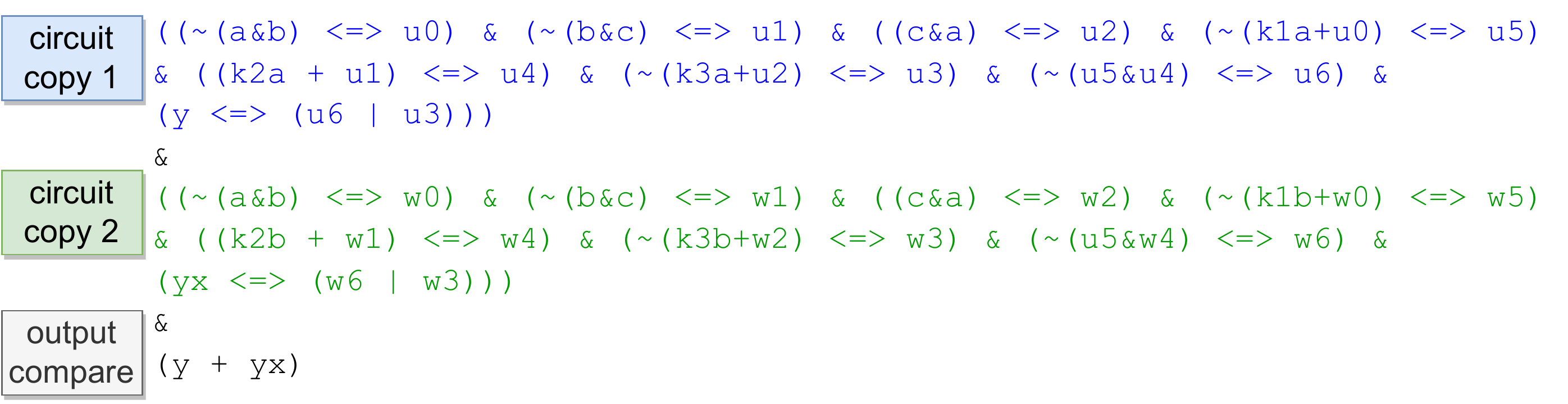}
    \vspace{-1em}
    \caption{Miter circuit as a formula: SAT Attack Iteration \#1}
    \label{fig:sat1}
\end{figure}

To begin, let us construct a miter circuit using two copies of the locked design. For readability, we will label the internal nets of the first copy as shown in \autoref{fig:lock-example} and represent the second copy's nets with ``w''. We distinguish between copy 1's keys and copy 2's keys by appending `a' and `b', respectively.  
The miter circuit is shown as a formula in \autoref{fig:sat1}.

\begin{figure}[h]
    \centering
    \includegraphics[width=0.9\textwidth]{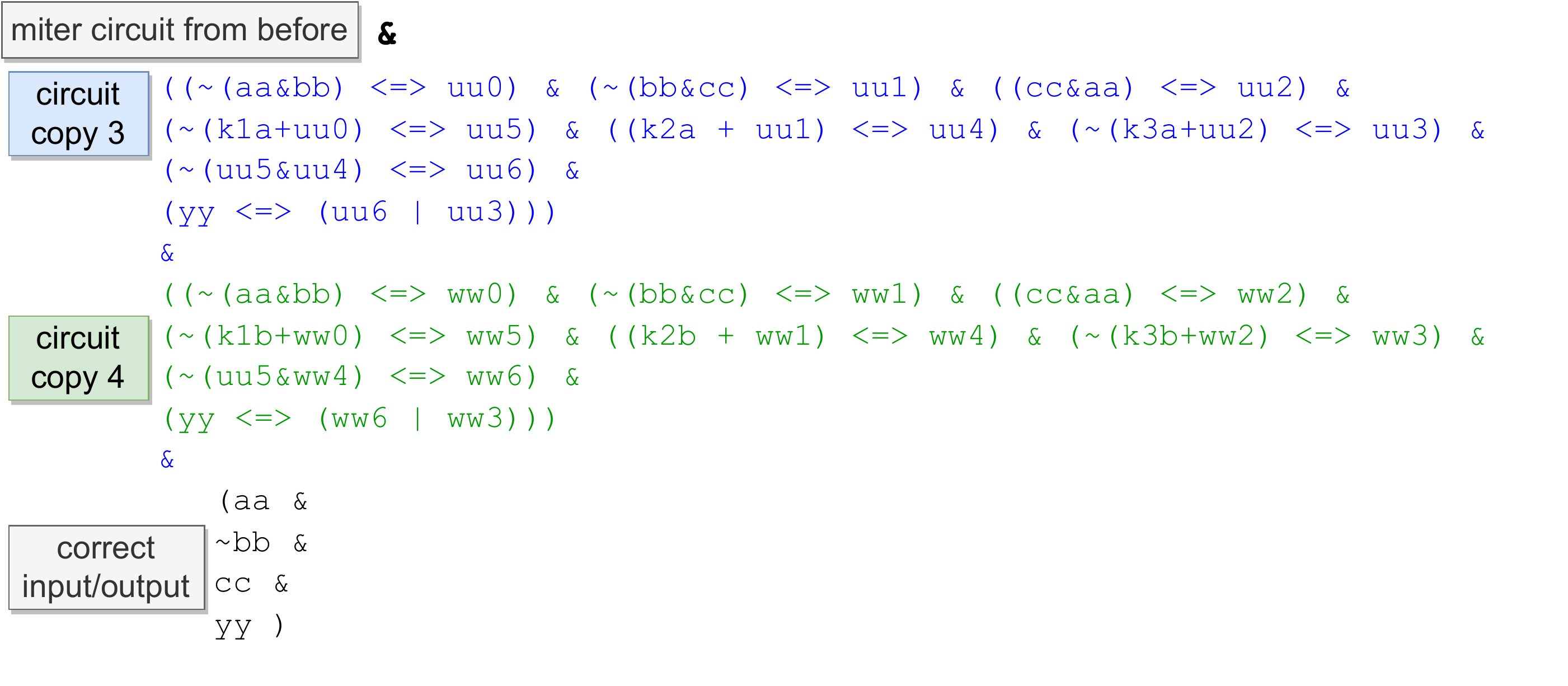}
    \vspace{-1em}
    \caption{SAT Attack Iteration \#2}
    \label{fig:sat2}
\end{figure}

When we solve using \texttt{dpll:old}, we receive the following feedback:

\begin{lstlisting}[breaklines=true,basicstyle=\small]
    Clause set is true if we assign values to variables as: u0 a -b u1 c u2 u5 k1a -u4 k2a u3 k3a u6 y w0 w1 w2 w5 k1b w4 
    -k2b -w3 -k3b -w6 -yx
\end{lstlisting}

In other words, the miter circuit formula is \textbf{satisfiable}! Of particular interest is the value of the inputs that the solver has found, being $a=1$, $b=0$ (see \texttt{-b} in the feedback), and $c=1$. This is a distinguishing input pattern. In this case, circuit copy 1 produces an output of $1$, while circuit copy 2 produces an output of $0$. 
We can check the Oracle to see which is the intended output; from \autoref{fig:lock-example}, the correct output should be $1$. 
As we can see in line 6 of \autoref{alg:sat_attack}, we should make a new formula by adding more copies of the circuit but with the constraint that the output produced with the DIP of $a=1,b=0,c=1$ should be $1$. 
The formula for the next iteration of the SAT attack thus looks like \autoref{fig:sat2}. 

\begin{figure}[h]
    \centering
    \includegraphics[width=0.9\textwidth]{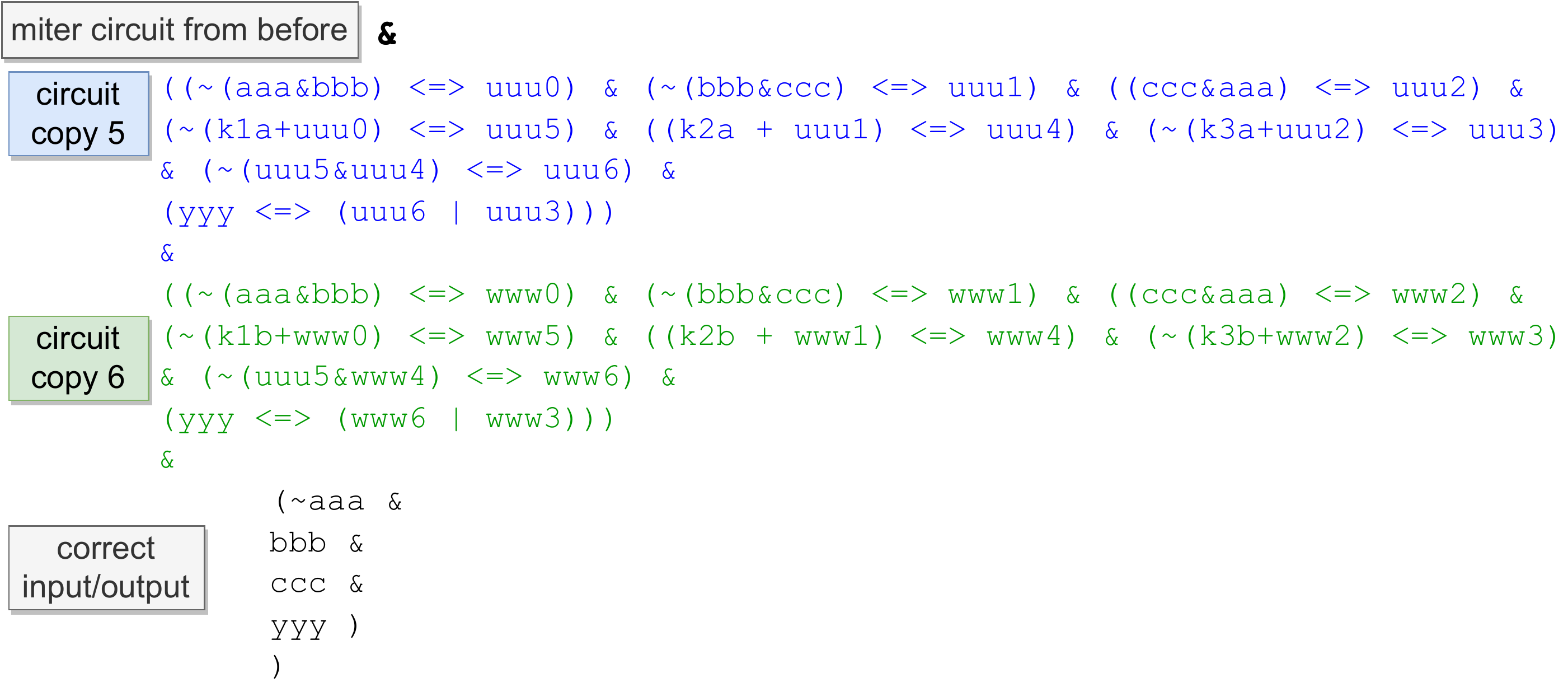}
    \vspace{-1em}
    \caption{SAT Attack Iteration \#3}
    \label{fig:sat3}
\end{figure}

We add the miter circuit formula from iteration \#1 to new copies of the circuit. Note the new internal node names ``uuX'' and ``wwX'' to represent the next copy, as well as ``aa'', ``bb'', ``cc'', and ``yy'' for the DIP and corresponding output from the oracle. 
Note also that the variables representing the key (e.g., k1a, k1b) are the same as those in the initial miter circuit formula; this forces keys found in future iterations to make the circuit produce the correct output for the DIP. 
Once again, using the SAT solver produces the following feedback: 

\begin{lstlisting}[breaklines=true,basicstyle=\small]
    Clause set is true if we assign values to variables as: u0 -a b -u1 c -u2 u5 k1a u4 k2a -u3 k3a -u6 -y w0 -w1 -w2 w5 
    k1b w4 k2b w3 -k3b -w6 yx uu0 aa -bb uu1 cc uu2 uu5 -uu4 uu3 uu6 yy ww0 ww1 ww2 ww5 -ww4 -ww3 ww6
\end{lstlisting}

This time, the DIP that is found is $a=0, b=1, c=1$, which the Oracles tells us should produce the output $1$. 
We can do yet another iteration of the SAT attack, adding \autoref{fig:sat3} to our growing formula. 

Running the SAT solver, we find that the formula is still satisifiable, with the following feedback: 

\begin{lstlisting}[breaklines=true,basicstyle=\small]
Clause set is true if we assign values to variables as: u0 -a b u1 -c -u2 u5 k1a -u4 k2a u3 -k3a u6 y w0 w1 -w2 w5 k1b w4 -k2b -w3 k3b -w6 -yx uu0 aa -bb uu1 cc uu2 uu5 -uu4 -uu3 uu6 yy ww0 ww1 ww2 ww5 ww4 ww3 -ww6 uuu0 -aaa bbb -uuu1 ccc -uuu2 uuu5 uuu4 uuu3 -uuu6 yyy www0 -www1 -www2 www5 -www4 -www3 www6
\end{lstlisting}

If we construct the new formula for another iteration and run that through the solver (see the Appendix for the full formula), we finally receive the following feedback:

\begin{lstlisting}[breaklines=true,basicstyle=\small]
Clause set is false for all possible assignments to variables.
\end{lstlisting}

The formula is \textbf{unsatisfiable}! This means that the solver can no longer find any DIPs; in other words, all the remaining inputs and keys will make the two copies of the circuit behave equivalently. 
Because the formula also adds constraints that any remaining keys make the circuit produce the correct input/output behavior (from the previous DIPs), any key that satisfies the formula as a whole, without the \texttt{y + yx} constraint, should be a correct key. Thus, running the formula with that constraint missing, gives us: 

\begin{lstlisting}[breaklines=true,basicstyle=\small]
Clause set is true if we assign values to variables as: -u0 a b 
-u1 c u2 -u5 k1a -u4 -k2a u3 k3a u6 y -w0 -w1 w2 -w5 k1b -w4 -k2b w3 k3b w6 yx uu0 aa -bb uu1 cc uu2 uu5 uu4 uu3 -uu6 yy 
ww0 ww1 ww2 ww5 ww4 ww3 -ww6 uuu0 -aaa bbb -uuu1 ccc -uuu2 uuu5 
-uuu4 -uuu3 uuu6 yyy www0 -www1 -www2 www5 -www4 -www3 www6 uuuu0 -aaaa bbbb uuuu1 -cccc -uuuu2 uuuu5 uuuu4 -uuuu3 -uuuu6 -yyyy wwww0 wwww1 -wwww2 wwww5 wwww4 -wwww3 -wwww6
\end{lstlisting}
If we hone in on the variables representing the key bits, we can see that $k1a = k1b = 1, k2a = k2b = 0, k3a = k3b = 1$.

\subsection{Discussion: What's next for logic locking?}

The arrival of the ``SAT attack'' marked a turning point in the logic locking domain. 
Recent survey works such as that by Chakraborty et al.~\cite{chakraborty_keynote_2020} provide a good overview of the various techniques which you should now be able to better appreciate. 
Several post-SAT techniques try to reduce the number of keys that are pruned with each iteration of the SAT attack (e.g., \cite{yasin_provably-secure_2017}), while others try to introduce circuit structures that cause issues for SAT solvers (e.g., \cite{shamsi_cyclic_2017}). 
New defense techniques are proposed and countered, even now, as the ``cat-and-mouse'' game continues. 
Other recent work includes the proposal of FPGA-based redaction~\cite{mohan_hardware_2021} or universal circuits~\cite{beerel_towards_2022}, among other strategies. 

While this tutorial covered the first formulation of the SAT attack from Subramanyan et al.~\cite{subramanyan_evaluating_2015}, there are a few more things we should note. 
As we mentioned in \autoref{sec:lock-tut}, this tutorial focused on a combinational design, so you are probably interested to know what happens in more realistic systems, where we have memory elements and sequential logic. 
In the early days of logic locking, the notion of Oracle access is often paired with the idea of a fully-scanned design with an adversary-accessible scan chain (as you explored in the previous case study in \autoref{sec:scan}). 
With design-for-test structures like the scan-chain, an adversary can treat the different parts of the design as combinational by scanning in input sequences and scanning out the register contents (assuming, of course, that the logic locking key is itself not trivially connected to the scan chain!). 
Thus, advances in scan chain protection, such as the recently proposed DisOrc~\cite{limaye_thwarting_2021}, provide an orthogonal means to protect against the SAT attack and other Oracle-based attacks on logic locking. 
Other approaches, like the KC2 attack~\cite{shamsi_kc2_2019}, incorporate the idea of \textit{sequential unrolling}, where the adversary makes several copies of the circuit in the Boolean formula to represent the inputs and outputs of the design across several clock cycles. 
However, as one can see, even with our simple tutorial walk-through, the Boolean formula grows very quickly!

Logic locking continues to evolve, with new defenses and attacks being proposed regularly by academia. 
Readers that are interested to explore more about logic locking can consider reading some of the following works. 
Setting aside Oracle-based attacks, there are several Oracle-less attacks as well. 
These include analyzing the circuit in terms of logic redundancy with incorrect keys~\cite{li_piercing_2019} as well as efforts focused on structural analysis and machine learning~\cite{chakraborty_sail_2021,sisejkovic_challenging_2021,alaql_scope_2021}. 
We direct interested readers to a useful survey paper on machine learning and logic locking by Sisejkovic et al.~\cite{sisejkovic_logic_2021}. 
Sequential obfuscation (e.g.,~\cite{chakraborty_harpoon_2009,rahman_practical_2022}), where the state space of a design is modified, is another emerging area. 

Recently, reconfigurable fabrics, such as embedded field-programmable gate arrays (eFPGAs), (e.g.,~\cite{mohan_hardware_2021,bhandari_exploring_2021,abideen_fpgas_2021,chowdhury_predictive_2022,shihab_design_2019}) have been proposed to withhold elements of the design. 
Working out what to ``redact'' or lock remains challenging, with several recent approaches emerging (e.g.,~\cite{tomajoli_alice_2022,chen_decoy_2020}).
For access to tools and benchmarks, we encourage the readers to check out \url{https://trust-hub.org/} and \url{https://cadforassurance.org/} -- these websites collate the results of several research efforts. 

Fundamentally, there is an ongoing need to formalize notions of security, trade-off security against the cost of each solution, and devise new means of attack and defense.

\section{Conclusions \label{sec:conclusions}}

In this paper, we provided a tutorial introduction to issues in hardware security. 
In particular, we presented two detailed case studies of problems in hardware security: attacking cryptography via the scan chain side channel and attacks on logic locking for hardware intellectual property protection. 
Through these pedagogical examples, we provide a foundation which readers can use to engage with more recent research in these domains. 
The tutorial examples are supported by an open access online resource hosted at \url{https://github.com/learn-hardware-security}. 

\begin{acks}
This research was supported in part by NSF Grant \#2039607. 
Any opinions, findings, and conclusions, or recommendations expressed are those of the author(s) and do not necessarily reflect the views of the National Science Foundation.

\end{acks}

\bibliographystyle{ACM-Reference-Format}
\bibliography{bib/ESWEEK_Tutorial}

\appendix

\section{DES Tables}
\label{sec:appendix:des-tables}
This appendix provides the various permutation tables used within DES~\cite{nist_data_1999}.
The general process for reading a permutation table is as follows.
The output bits are generated in-order using the input, with the input indexed using the appropriate value from the table.
For example, the first (left-most / MSB) bit of the output of $IP$ will be the value of the 58th bit of the input (refer to \autoref{tbl:des-IP}).
Then, the second bit will be the 50th bit of the input, and so on.
The IP, FP, E, P, PC1, and PC2 tables are presented as tables only for ease of presentation. They are vectors, not tables.

\subsection{Encryption / Decryption permutation tables}

\begin{table}[h]
\caption{Initial Permutation (IP)}
\label{tbl:des-IP}
\begin{tabular}{|r|r|r|r|r|r|r|r|}
\hline
58 & 50 & 42 & 34 & 26 & 18 & 10 & 2 \\ \hline
60 & 52 & 44 & 36 & 28 & 20 & 12 & 4 \\ \hline
62 & 54 & 46 & 38 & 30 & 22 & 14 & 6 \\ \hline
64 & 56 & 48 & 40 & 32 & 24 & 16 & 8 \\ \hline
57 & 49 & 41 & 33 & 25 & 17 & 9  & 1 \\ \hline
59 & 51 & 43 & 35 & 27 & 19 & 11 & 3 \\ \hline
61 & 53 & 45 & 37 & 29 & 21 & 13 & 5 \\ \hline
63 & 55 & 47 & 39 & 31 & 23 & 15 & 7 \\ \hline
\end{tabular}
\end{table}

\begin{table}[H]
\caption{Final Permutation (FP) (is equal to the inverse of the IP table, IP$^{-1}$).}
\label{tbl:des-FP}
\begin{tabular}{|r|r|r|r|r|r|r|r|}
\hline
40 & 8 & 48 & 16 & 56 & 24 & 64 & 32 \\ \hline
39 & 7 & 47 & 15 & 55 & 23 & 63 & 31 \\ \hline
38 & 6 & 46 & 14 & 54 & 22 & 62 & 30 \\ \hline
37 & 5 & 45 & 13 & 53 & 21 & 61 & 29 \\ \hline
36 & 4 & 44 & 12 & 52 & 20 & 60 & 28 \\ \hline
35 & 3 & 43 & 11 & 51 & 19 & 59 & 27 \\ \hline
34 & 2 & 42 & 10 & 50 & 18 & 58 & 26 \\ \hline
33 & 1 & 41 & 9  & 49 & 17 & 57 & 25 \\ \hline
\end{tabular}
\end{table}

\begin{table}[H]
\caption{Expansion Function ($E$). Note the repeating bits (e.g. 32, 1).}
\label{tbl:des-E}
\begin{tabular}{|r|r|r|r|r|r|}
\hline
32 & 1  & 2  & 3  & 4  & 5  \\ \hline
4  & 5  & 6  & 7  & 8  & 9  \\ \hline
8  & 9  & 10 & 11 & 12 & 13 \\ \hline
12 & 13 & 14 & 15 & 16 & 17 \\ \hline
16 & 17 & 18 & 19 & 20 & 21 \\ \hline
20 & 21 & 22 & 23 & 24 & 25 \\ \hline
24 & 25 & 26 & 27 & 28 & 29 \\ \hline
28 & 29 & 30 & 31 & 32 & 1  \\ \hline
\end{tabular}
\end{table}

\begin{table}[H]
\caption{Permutation ($P$).}
\label{tbl:des-P}
\begin{tabular}{|r|r|r|r|r|r|r|r|}
\hline
16 & 7  & 20 & 21 & 29 & 12 & 28 & 17 \\ \hline
1  & 15 & 23 & 26 & 5  & 18 & 31 & 10 \\ \hline
2  & 8  & 24 & 14 & 32 & 27 & 3  & 9  \\ \hline
19 & 13 & 30 & 6  & 22 & 11 & 4  & 25 \\ \hline
\end{tabular}
\end{table}

\begin{table}[H]
\caption{DES s-boxes}
\label{tbl:des-sbox-all}
\resizebox{\linewidth}{!}{%
\begin{tabular}{rrrrrrrrrrrrrrrrr}
\hline
\multicolumn{1}{|r|}{\textbf{S1}}     & \multicolumn{1}{r|}{\textbf{x0000x}} & \multicolumn{1}{r|}{\textbf{x0001x}} & \multicolumn{1}{r|}{\textbf{x0010x}} & \multicolumn{1}{r|}{\textbf{x0011x}} & \multicolumn{1}{r|}{\textbf{x0100x}} & \multicolumn{1}{r|}{\textbf{x0101x}} & \multicolumn{1}{r|}{\textbf{x0110x}} & \multicolumn{1}{r|}{\textbf{x0111x}} & \multicolumn{1}{r|}{\textbf{x1000x}} & \multicolumn{1}{r|}{\textbf{x1001x}} & \multicolumn{1}{r|}{\textbf{x1010x}} & \multicolumn{1}{r|}{\textbf{x1011x}} & \multicolumn{1}{r|}{\textbf{x1100x}} & \multicolumn{1}{r|}{\textbf{x1101x}} & \multicolumn{1}{r|}{\textbf{x1110x}} & \multicolumn{1}{r|}{\textbf{x1111x}} \\ \hline
\multicolumn{1}{|r|}{\textbf{0yyyy0}} & \multicolumn{1}{r|}{14}              & \multicolumn{1}{r|}{4}               & \multicolumn{1}{r|}{13}              & \multicolumn{1}{r|}{1}               & \multicolumn{1}{r|}{2}               & \multicolumn{1}{r|}{15}              & \multicolumn{1}{r|}{11}              & \multicolumn{1}{r|}{8}               & \multicolumn{1}{r|}{3}               & \multicolumn{1}{r|}{10}              & \multicolumn{1}{r|}{6}               & \multicolumn{1}{r|}{12}              & \multicolumn{1}{r|}{5}               & \multicolumn{1}{r|}{9}               & \multicolumn{1}{r|}{0}               & \multicolumn{1}{r|}{7}               \\ \hline
\multicolumn{1}{|r|}{\textbf{0yyyy1}} & \multicolumn{1}{r|}{0}               & \multicolumn{1}{r|}{15}              & \multicolumn{1}{r|}{7}               & \multicolumn{1}{r|}{4}               & \multicolumn{1}{r|}{14}              & \multicolumn{1}{r|}{2}               & \multicolumn{1}{r|}{13}              & \multicolumn{1}{r|}{1}               & \multicolumn{1}{r|}{10}              & \multicolumn{1}{r|}{6}               & \multicolumn{1}{r|}{12}              & \multicolumn{1}{r|}{11}              & \multicolumn{1}{r|}{9}               & \multicolumn{1}{r|}{5}               & \multicolumn{1}{r|}{3}               & \multicolumn{1}{r|}{8}               \\ \hline
\multicolumn{1}{|r|}{\textbf{1yyyy0}} & \multicolumn{1}{r|}{4}               & \multicolumn{1}{r|}{1}               & \multicolumn{1}{r|}{14}              & \multicolumn{1}{r|}{8}               & \multicolumn{1}{r|}{13}              & \multicolumn{1}{r|}{6}               & \multicolumn{1}{r|}{2}               & \multicolumn{1}{r|}{11}              & \multicolumn{1}{r|}{15}              & \multicolumn{1}{r|}{12}              & \multicolumn{1}{r|}{9}               & \multicolumn{1}{r|}{7}               & \multicolumn{1}{r|}{3}               & \multicolumn{1}{r|}{10}              & \multicolumn{1}{r|}{5}               & \multicolumn{1}{r|}{0}               \\ \hline
\multicolumn{1}{|r|}{\textbf{1yyyy1}} & \multicolumn{1}{r|}{15}              & \multicolumn{1}{r|}{12}              & \multicolumn{1}{r|}{8}               & \multicolumn{1}{r|}{2}               & \multicolumn{1}{r|}{4}               & \multicolumn{1}{r|}{9}               & \multicolumn{1}{r|}{1}               & \multicolumn{1}{r|}{7}               & \multicolumn{1}{r|}{5}               & \multicolumn{1}{r|}{11}              & \multicolumn{1}{r|}{3}               & \multicolumn{1}{r|}{14}              & \multicolumn{1}{r|}{10}              & \multicolumn{1}{r|}{0}               & \multicolumn{1}{r|}{6}               & \multicolumn{1}{r|}{13}              \\ \hline
\multicolumn{17}{c}{\textbf{}}                                                                                                                                                                                                                                                                                                                                                                                                                                                                                                                                                                                                                                                        \\ \hline
\multicolumn{1}{|r|}{\textbf{S2}}     & \multicolumn{1}{r|}{\textbf{x0000x}} & \multicolumn{1}{r|}{\textbf{x0001x}} & \multicolumn{1}{r|}{\textbf{x0010x}} & \multicolumn{1}{r|}{\textbf{x0011x}} & \multicolumn{1}{r|}{\textbf{x0100x}} & \multicolumn{1}{r|}{\textbf{x0101x}} & \multicolumn{1}{r|}{\textbf{x0110x}} & \multicolumn{1}{r|}{\textbf{x0111x}} & \multicolumn{1}{r|}{\textbf{x1000x}} & \multicolumn{1}{r|}{\textbf{x1001x}} & \multicolumn{1}{r|}{\textbf{x1010x}} & \multicolumn{1}{r|}{\textbf{x1011x}} & \multicolumn{1}{r|}{\textbf{x1100x}} & \multicolumn{1}{r|}{\textbf{x1101x}} & \multicolumn{1}{r|}{\textbf{x1110x}} & \multicolumn{1}{r|}{\textbf{x1111x}} \\ \hline
\multicolumn{1}{|r|}{\textbf{0yyyy0}} & \multicolumn{1}{r|}{15}              & \multicolumn{1}{r|}{1}               & \multicolumn{1}{r|}{8}               & \multicolumn{1}{r|}{14}              & \multicolumn{1}{r|}{6}               & \multicolumn{1}{r|}{11}              & \multicolumn{1}{r|}{3}               & \multicolumn{1}{r|}{4}               & \multicolumn{1}{r|}{9}               & \multicolumn{1}{r|}{7}               & \multicolumn{1}{r|}{2}               & \multicolumn{1}{r|}{13}              & \multicolumn{1}{r|}{12}              & \multicolumn{1}{r|}{0}               & \multicolumn{1}{r|}{5}               & \multicolumn{1}{r|}{10}              \\ \hline
\multicolumn{1}{|r|}{\textbf{0yyyy1}} & \multicolumn{1}{r|}{3}               & \multicolumn{1}{r|}{13}              & \multicolumn{1}{r|}{4}               & \multicolumn{1}{r|}{7}               & \multicolumn{1}{r|}{15}              & \multicolumn{1}{r|}{2}               & \multicolumn{1}{r|}{8}               & \multicolumn{1}{r|}{14}              & \multicolumn{1}{r|}{12}              & \multicolumn{1}{r|}{0}               & \multicolumn{1}{r|}{1}               & \multicolumn{1}{r|}{10}              & \multicolumn{1}{r|}{6}               & \multicolumn{1}{r|}{9}               & \multicolumn{1}{r|}{11}              & \multicolumn{1}{r|}{5}               \\ \hline
\multicolumn{1}{|r|}{\textbf{1yyyy0}} & \multicolumn{1}{r|}{0}               & \multicolumn{1}{r|}{14}              & \multicolumn{1}{r|}{7}               & \multicolumn{1}{r|}{11}              & \multicolumn{1}{r|}{10}              & \multicolumn{1}{r|}{4}               & \multicolumn{1}{r|}{13}              & \multicolumn{1}{r|}{1}               & \multicolumn{1}{r|}{5}               & \multicolumn{1}{r|}{8}               & \multicolumn{1}{r|}{12}              & \multicolumn{1}{r|}{6}               & \multicolumn{1}{r|}{9}               & \multicolumn{1}{r|}{3}               & \multicolumn{1}{r|}{2}               & \multicolumn{1}{r|}{15}              \\ \hline
\multicolumn{1}{|r|}{\textbf{1yyyy1}} & \multicolumn{1}{r|}{13}              & \multicolumn{1}{r|}{8}               & \multicolumn{1}{r|}{10}              & \multicolumn{1}{r|}{1}               & \multicolumn{1}{r|}{3}               & \multicolumn{1}{r|}{15}              & \multicolumn{1}{r|}{4}               & \multicolumn{1}{r|}{2}               & \multicolumn{1}{r|}{11}              & \multicolumn{1}{r|}{6}               & \multicolumn{1}{r|}{7}               & \multicolumn{1}{r|}{12}              & \multicolumn{1}{r|}{0}               & \multicolumn{1}{r|}{5}               & \multicolumn{1}{r|}{14}              & \multicolumn{1}{r|}{9}               \\ \hline
\multicolumn{17}{c}{\textbf{}}                                                                                                                                                                                                                                                                                                                                                                                                                                                                                                                                                                                                                                                        \\ \hline
\multicolumn{1}{|r|}{\textbf{S3}}     & \multicolumn{1}{r|}{\textbf{x0000x}} & \multicolumn{1}{r|}{\textbf{x0001x}} & \multicolumn{1}{r|}{\textbf{x0010x}} & \multicolumn{1}{r|}{\textbf{x0011x}} & \multicolumn{1}{r|}{\textbf{x0100x}} & \multicolumn{1}{r|}{\textbf{x0101x}} & \multicolumn{1}{r|}{\textbf{x0110x}} & \multicolumn{1}{r|}{\textbf{x0111x}} & \multicolumn{1}{r|}{\textbf{x1000x}} & \multicolumn{1}{r|}{\textbf{x1001x}} & \multicolumn{1}{r|}{\textbf{x1010x}} & \multicolumn{1}{r|}{\textbf{x1011x}} & \multicolumn{1}{r|}{\textbf{x1100x}} & \multicolumn{1}{r|}{\textbf{x1101x}} & \multicolumn{1}{r|}{\textbf{x1110x}} & \multicolumn{1}{r|}{\textbf{x1111x}} \\ \hline
\multicolumn{1}{|r|}{\textbf{0yyyy0}} & \multicolumn{1}{r|}{10}              & \multicolumn{1}{r|}{0}               & \multicolumn{1}{r|}{9}               & \multicolumn{1}{r|}{14}              & \multicolumn{1}{r|}{6}               & \multicolumn{1}{r|}{3}               & \multicolumn{1}{r|}{15}              & \multicolumn{1}{r|}{5}               & \multicolumn{1}{r|}{1}               & \multicolumn{1}{r|}{13}              & \multicolumn{1}{r|}{12}              & \multicolumn{1}{r|}{7}               & \multicolumn{1}{r|}{11}              & \multicolumn{1}{r|}{4}               & \multicolumn{1}{r|}{2}               & \multicolumn{1}{r|}{8}               \\ \hline
\multicolumn{1}{|r|}{\textbf{0yyyy1}} & \multicolumn{1}{r|}{13}              & \multicolumn{1}{r|}{7}               & \multicolumn{1}{r|}{0}               & \multicolumn{1}{r|}{9}               & \multicolumn{1}{r|}{3}               & \multicolumn{1}{r|}{4}               & \multicolumn{1}{r|}{6}               & \multicolumn{1}{r|}{10}              & \multicolumn{1}{r|}{2}               & \multicolumn{1}{r|}{8}               & \multicolumn{1}{r|}{5}               & \multicolumn{1}{r|}{14}              & \multicolumn{1}{r|}{12}              & \multicolumn{1}{r|}{11}              & \multicolumn{1}{r|}{15}              & \multicolumn{1}{r|}{1}               \\ \hline
\multicolumn{1}{|r|}{\textbf{1yyyy0}} & \multicolumn{1}{r|}{13}              & \multicolumn{1}{r|}{6}               & \multicolumn{1}{r|}{4}               & \multicolumn{1}{r|}{9}               & \multicolumn{1}{r|}{8}               & \multicolumn{1}{r|}{15}              & \multicolumn{1}{r|}{3}               & \multicolumn{1}{r|}{0}               & \multicolumn{1}{r|}{11}              & \multicolumn{1}{r|}{1}               & \multicolumn{1}{r|}{2}               & \multicolumn{1}{r|}{12}              & \multicolumn{1}{r|}{5}               & \multicolumn{1}{r|}{10}              & \multicolumn{1}{r|}{14}              & \multicolumn{1}{r|}{7}               \\ \hline
\multicolumn{1}{|r|}{\textbf{1yyyy1}} & \multicolumn{1}{r|}{1}               & \multicolumn{1}{r|}{10}              & \multicolumn{1}{r|}{13}              & \multicolumn{1}{r|}{0}               & \multicolumn{1}{r|}{6}               & \multicolumn{1}{r|}{9}               & \multicolumn{1}{r|}{8}               & \multicolumn{1}{r|}{7}               & \multicolumn{1}{r|}{4}               & \multicolumn{1}{r|}{15}              & \multicolumn{1}{r|}{14}              & \multicolumn{1}{r|}{3}               & \multicolumn{1}{r|}{11}              & \multicolumn{1}{r|}{5}               & \multicolumn{1}{r|}{2}               & \multicolumn{1}{r|}{12}              \\ \hline
\multicolumn{17}{c}{\textbf{}}                                                                                                                                                                                                                                                                                                                                                                                                                                                                                                                                                                                                                                                        \\ \hline
\multicolumn{1}{|r|}{\textbf{S4}}     & \multicolumn{1}{r|}{\textbf{x0000x}} & \multicolumn{1}{r|}{\textbf{x0001x}} & \multicolumn{1}{r|}{\textbf{x0010x}} & \multicolumn{1}{r|}{\textbf{x0011x}} & \multicolumn{1}{r|}{\textbf{x0100x}} & \multicolumn{1}{r|}{\textbf{x0101x}} & \multicolumn{1}{r|}{\textbf{x0110x}} & \multicolumn{1}{r|}{\textbf{x0111x}} & \multicolumn{1}{r|}{\textbf{x1000x}} & \multicolumn{1}{r|}{\textbf{x1001x}} & \multicolumn{1}{r|}{\textbf{x1010x}} & \multicolumn{1}{r|}{\textbf{x1011x}} & \multicolumn{1}{r|}{\textbf{x1100x}} & \multicolumn{1}{r|}{\textbf{x1101x}} & \multicolumn{1}{r|}{\textbf{x1110x}} & \multicolumn{1}{r|}{\textbf{x1111x}} \\ \hline
\multicolumn{1}{|r|}{\textbf{0yyyy0}} & \multicolumn{1}{r|}{7}               & \multicolumn{1}{r|}{13}              & \multicolumn{1}{r|}{14}              & \multicolumn{1}{r|}{3}               & \multicolumn{1}{r|}{0}               & \multicolumn{1}{r|}{6}               & \multicolumn{1}{r|}{9}               & \multicolumn{1}{r|}{10}              & \multicolumn{1}{r|}{1}               & \multicolumn{1}{r|}{2}               & \multicolumn{1}{r|}{8}               & \multicolumn{1}{r|}{5}               & \multicolumn{1}{r|}{11}              & \multicolumn{1}{r|}{12}              & \multicolumn{1}{r|}{4}               & \multicolumn{1}{r|}{15}              \\ \hline
\multicolumn{1}{|r|}{\textbf{0yyyy1}} & \multicolumn{1}{r|}{13}              & \multicolumn{1}{r|}{8}               & \multicolumn{1}{r|}{11}              & \multicolumn{1}{r|}{5}               & \multicolumn{1}{r|}{6}               & \multicolumn{1}{r|}{15}              & \multicolumn{1}{r|}{0}               & \multicolumn{1}{r|}{3}               & \multicolumn{1}{r|}{4}               & \multicolumn{1}{r|}{7}               & \multicolumn{1}{r|}{2}               & \multicolumn{1}{r|}{12}              & \multicolumn{1}{r|}{1}               & \multicolumn{1}{r|}{10}              & \multicolumn{1}{r|}{14}              & \multicolumn{1}{r|}{9}               \\ \hline
\multicolumn{1}{|r|}{\textbf{1yyyy0}} & \multicolumn{1}{r|}{10}              & \multicolumn{1}{r|}{6}               & \multicolumn{1}{r|}{9}               & \multicolumn{1}{r|}{0}               & \multicolumn{1}{r|}{12}              & \multicolumn{1}{r|}{11}              & \multicolumn{1}{r|}{7}               & \multicolumn{1}{r|}{13}              & \multicolumn{1}{r|}{15}              & \multicolumn{1}{r|}{1}               & \multicolumn{1}{r|}{3}               & \multicolumn{1}{r|}{14}              & \multicolumn{1}{r|}{5}               & \multicolumn{1}{r|}{2}               & \multicolumn{1}{r|}{8}               & \multicolumn{1}{r|}{4}               \\ \hline
\multicolumn{1}{|r|}{\textbf{1yyyy1}} & \multicolumn{1}{r|}{3}               & \multicolumn{1}{r|}{15}              & \multicolumn{1}{r|}{0}               & \multicolumn{1}{r|}{6}               & \multicolumn{1}{r|}{10}              & \multicolumn{1}{r|}{1}               & \multicolumn{1}{r|}{13}              & \multicolumn{1}{r|}{8}               & \multicolumn{1}{r|}{9}               & \multicolumn{1}{r|}{4}               & \multicolumn{1}{r|}{5}               & \multicolumn{1}{r|}{11}              & \multicolumn{1}{r|}{12}              & \multicolumn{1}{r|}{7}               & \multicolumn{1}{r|}{2}               & \multicolumn{1}{r|}{14}              \\ \hline
\multicolumn{17}{c}{\textbf{}}                                                                                                                                                                                                                                                                                                                                                                                                                                                                                                                                                                                                                                                        \\ \hline
\multicolumn{1}{|r|}{\textbf{S5}}     & \multicolumn{1}{r|}{\textbf{x0000x}} & \multicolumn{1}{r|}{\textbf{x0001x}} & \multicolumn{1}{r|}{\textbf{x0010x}} & \multicolumn{1}{r|}{\textbf{x0011x}} & \multicolumn{1}{r|}{\textbf{x0100x}} & \multicolumn{1}{r|}{\textbf{x0101x}} & \multicolumn{1}{r|}{\textbf{x0110x}} & \multicolumn{1}{r|}{\textbf{x0111x}} & \multicolumn{1}{r|}{\textbf{x1000x}} & \multicolumn{1}{r|}{\textbf{x1001x}} & \multicolumn{1}{r|}{\textbf{x1010x}} & \multicolumn{1}{r|}{\textbf{x1011x}} & \multicolumn{1}{r|}{\textbf{x1100x}} & \multicolumn{1}{r|}{\textbf{x1101x}} & \multicolumn{1}{r|}{\textbf{x1110x}} & \multicolumn{1}{r|}{\textbf{x1111x}} \\ \hline
\multicolumn{1}{|r|}{\textbf{0yyyy0}} & \multicolumn{1}{r|}{2}               & \multicolumn{1}{r|}{12}              & \multicolumn{1}{r|}{4}               & \multicolumn{1}{r|}{1}               & \multicolumn{1}{r|}{7}               & \multicolumn{1}{r|}{10}              & \multicolumn{1}{r|}{11}              & \multicolumn{1}{r|}{6}               & \multicolumn{1}{r|}{8}               & \multicolumn{1}{r|}{5}               & \multicolumn{1}{r|}{3}               & \multicolumn{1}{r|}{15}              & \multicolumn{1}{r|}{13}              & \multicolumn{1}{r|}{0}               & \multicolumn{1}{r|}{14}              & \multicolumn{1}{r|}{9}               \\ \hline
\multicolumn{1}{|r|}{\textbf{0yyyy1}} & \multicolumn{1}{r|}{14}              & \multicolumn{1}{r|}{11}              & \multicolumn{1}{r|}{2}               & \multicolumn{1}{r|}{12}              & \multicolumn{1}{r|}{4}               & \multicolumn{1}{r|}{7}               & \multicolumn{1}{r|}{13}              & \multicolumn{1}{r|}{1}               & \multicolumn{1}{r|}{5}               & \multicolumn{1}{r|}{0}               & \multicolumn{1}{r|}{15}              & \multicolumn{1}{r|}{10}              & \multicolumn{1}{r|}{3}               & \multicolumn{1}{r|}{9}               & \multicolumn{1}{r|}{8}               & \multicolumn{1}{r|}{6}               \\ \hline
\multicolumn{1}{|r|}{\textbf{1yyyy0}} & \multicolumn{1}{r|}{4}               & \multicolumn{1}{r|}{2}               & \multicolumn{1}{r|}{1}               & \multicolumn{1}{r|}{11}              & \multicolumn{1}{r|}{10}              & \multicolumn{1}{r|}{13}              & \multicolumn{1}{r|}{7}               & \multicolumn{1}{r|}{8}               & \multicolumn{1}{r|}{15}              & \multicolumn{1}{r|}{9}               & \multicolumn{1}{r|}{12}              & \multicolumn{1}{r|}{5}               & \multicolumn{1}{r|}{6}               & \multicolumn{1}{r|}{3}               & \multicolumn{1}{r|}{0}               & \multicolumn{1}{r|}{14}              \\ \hline
\multicolumn{1}{|r|}{\textbf{1yyyy1}} & \multicolumn{1}{r|}{11}              & \multicolumn{1}{r|}{8}               & \multicolumn{1}{r|}{12}              & \multicolumn{1}{r|}{7}               & \multicolumn{1}{r|}{1}               & \multicolumn{1}{r|}{14}              & \multicolumn{1}{r|}{2}               & \multicolumn{1}{r|}{13}              & \multicolumn{1}{r|}{6}               & \multicolumn{1}{r|}{15}              & \multicolumn{1}{r|}{0}               & \multicolumn{1}{r|}{9}               & \multicolumn{1}{r|}{10}              & \multicolumn{1}{r|}{4}               & \multicolumn{1}{r|}{5}               & \multicolumn{1}{r|}{3}               \\ \hline
\multicolumn{17}{c}{\textbf{}}                                                                                                                                                                                                                                                                                                                                                                                                                                                                                                                                                                                                                                                        \\ \hline
\multicolumn{1}{|r|}{\textbf{S6}}     & \multicolumn{1}{r|}{\textbf{x0000x}} & \multicolumn{1}{r|}{\textbf{x0001x}} & \multicolumn{1}{r|}{\textbf{x0010x}} & \multicolumn{1}{r|}{\textbf{x0011x}} & \multicolumn{1}{r|}{\textbf{x0100x}} & \multicolumn{1}{r|}{\textbf{x0101x}} & \multicolumn{1}{r|}{\textbf{x0110x}} & \multicolumn{1}{r|}{\textbf{x0111x}} & \multicolumn{1}{r|}{\textbf{x1000x}} & \multicolumn{1}{r|}{\textbf{x1001x}} & \multicolumn{1}{r|}{\textbf{x1010x}} & \multicolumn{1}{r|}{\textbf{x1011x}} & \multicolumn{1}{r|}{\textbf{x1100x}} & \multicolumn{1}{r|}{\textbf{x1101x}} & \multicolumn{1}{r|}{\textbf{x1110x}} & \multicolumn{1}{r|}{\textbf{x1111x}} \\ \hline
\multicolumn{1}{|r|}{\textbf{0yyyy0}} & \multicolumn{1}{r|}{12}              & \multicolumn{1}{r|}{1}               & \multicolumn{1}{r|}{10}              & \multicolumn{1}{r|}{15}              & \multicolumn{1}{r|}{9}               & \multicolumn{1}{r|}{2}               & \multicolumn{1}{r|}{6}               & \multicolumn{1}{r|}{8}               & \multicolumn{1}{r|}{0}               & \multicolumn{1}{r|}{13}              & \multicolumn{1}{r|}{3}               & \multicolumn{1}{r|}{4}               & \multicolumn{1}{r|}{14}              & \multicolumn{1}{r|}{7}               & \multicolumn{1}{r|}{5}               & \multicolumn{1}{r|}{11}              \\ \hline
\multicolumn{1}{|r|}{\textbf{0yyyy1}} & \multicolumn{1}{r|}{10}              & \multicolumn{1}{r|}{15}              & \multicolumn{1}{r|}{4}               & \multicolumn{1}{r|}{2}               & \multicolumn{1}{r|}{7}               & \multicolumn{1}{r|}{12}              & \multicolumn{1}{r|}{9}               & \multicolumn{1}{r|}{5}               & \multicolumn{1}{r|}{6}               & \multicolumn{1}{r|}{1}               & \multicolumn{1}{r|}{13}              & \multicolumn{1}{r|}{14}              & \multicolumn{1}{r|}{0}               & \multicolumn{1}{r|}{11}              & \multicolumn{1}{r|}{3}               & \multicolumn{1}{r|}{8}               \\ \hline
\multicolumn{1}{|r|}{\textbf{1yyyy0}} & \multicolumn{1}{r|}{9}               & \multicolumn{1}{r|}{14}              & \multicolumn{1}{r|}{15}              & \multicolumn{1}{r|}{5}               & \multicolumn{1}{r|}{2}               & \multicolumn{1}{r|}{8}               & \multicolumn{1}{r|}{12}              & \multicolumn{1}{r|}{3}               & \multicolumn{1}{r|}{7}               & \multicolumn{1}{r|}{0}               & \multicolumn{1}{r|}{4}               & \multicolumn{1}{r|}{10}              & \multicolumn{1}{r|}{1}               & \multicolumn{1}{r|}{13}              & \multicolumn{1}{r|}{11}              & \multicolumn{1}{r|}{6}               \\ \hline
\multicolumn{1}{|r|}{\textbf{1yyyy1}} & \multicolumn{1}{r|}{4}               & \multicolumn{1}{r|}{3}               & \multicolumn{1}{r|}{2}               & \multicolumn{1}{r|}{12}              & \multicolumn{1}{r|}{9}               & \multicolumn{1}{r|}{5}               & \multicolumn{1}{r|}{15}              & \multicolumn{1}{r|}{10}              & \multicolumn{1}{r|}{11}              & \multicolumn{1}{r|}{14}              & \multicolumn{1}{r|}{1}               & \multicolumn{1}{r|}{7}               & \multicolumn{1}{r|}{6}               & \multicolumn{1}{r|}{0}               & \multicolumn{1}{r|}{8}               & \multicolumn{1}{r|}{13}              \\ \hline
\multicolumn{17}{c}{\textbf{}}                                                                                                                                                                                                                                                                                                                                                                                                                                                                                                                                                                                                                                                        \\ \hline
\multicolumn{1}{|r|}{\textbf{S7}}     & \multicolumn{1}{r|}{\textbf{x0000x}} & \multicolumn{1}{r|}{\textbf{x0001x}} & \multicolumn{1}{r|}{\textbf{x0010x}} & \multicolumn{1}{r|}{\textbf{x0011x}} & \multicolumn{1}{r|}{\textbf{x0100x}} & \multicolumn{1}{r|}{\textbf{x0101x}} & \multicolumn{1}{r|}{\textbf{x0110x}} & \multicolumn{1}{r|}{\textbf{x0111x}} & \multicolumn{1}{r|}{\textbf{x1000x}} & \multicolumn{1}{r|}{\textbf{x1001x}} & \multicolumn{1}{r|}{\textbf{x1010x}} & \multicolumn{1}{r|}{\textbf{x1011x}} & \multicolumn{1}{r|}{\textbf{x1100x}} & \multicolumn{1}{r|}{\textbf{x1101x}} & \multicolumn{1}{r|}{\textbf{x1110x}} & \multicolumn{1}{r|}{\textbf{x1111x}} \\ \hline
\multicolumn{1}{|r|}{\textbf{0yyyy0}} & \multicolumn{1}{r|}{4}               & \multicolumn{1}{r|}{11}              & \multicolumn{1}{r|}{2}               & \multicolumn{1}{r|}{14}              & \multicolumn{1}{r|}{15}              & \multicolumn{1}{r|}{0}               & \multicolumn{1}{r|}{8}               & \multicolumn{1}{r|}{13}              & \multicolumn{1}{r|}{3}               & \multicolumn{1}{r|}{12}              & \multicolumn{1}{r|}{9}               & \multicolumn{1}{r|}{7}               & \multicolumn{1}{r|}{5}               & \multicolumn{1}{r|}{10}              & \multicolumn{1}{r|}{6}               & \multicolumn{1}{r|}{1}               \\ \hline
\multicolumn{1}{|r|}{\textbf{0yyyy1}} & \multicolumn{1}{r|}{13}              & \multicolumn{1}{r|}{0}               & \multicolumn{1}{r|}{11}              & \multicolumn{1}{r|}{7}               & \multicolumn{1}{r|}{4}               & \multicolumn{1}{r|}{9}               & \multicolumn{1}{r|}{1}               & \multicolumn{1}{r|}{10}              & \multicolumn{1}{r|}{14}              & \multicolumn{1}{r|}{3}               & \multicolumn{1}{r|}{5}               & \multicolumn{1}{r|}{12}              & \multicolumn{1}{r|}{2}               & \multicolumn{1}{r|}{15}              & \multicolumn{1}{r|}{8}               & \multicolumn{1}{r|}{6}               \\ \hline
\multicolumn{1}{|r|}{\textbf{1yyyy0}} & \multicolumn{1}{r|}{1}               & \multicolumn{1}{r|}{4}               & \multicolumn{1}{r|}{11}              & \multicolumn{1}{r|}{13}              & \multicolumn{1}{r|}{12}              & \multicolumn{1}{r|}{3}               & \multicolumn{1}{r|}{7}               & \multicolumn{1}{r|}{14}              & \multicolumn{1}{r|}{10}              & \multicolumn{1}{r|}{15}              & \multicolumn{1}{r|}{6}               & \multicolumn{1}{r|}{8}               & \multicolumn{1}{r|}{0}               & \multicolumn{1}{r|}{5}               & \multicolumn{1}{r|}{9}               & \multicolumn{1}{r|}{2}               \\ \hline
\multicolumn{1}{|r|}{\textbf{1yyyy1}} & \multicolumn{1}{r|}{6}               & \multicolumn{1}{r|}{11}              & \multicolumn{1}{r|}{13}              & \multicolumn{1}{r|}{8}               & \multicolumn{1}{r|}{1}               & \multicolumn{1}{r|}{4}               & \multicolumn{1}{r|}{10}              & \multicolumn{1}{r|}{7}               & \multicolumn{1}{r|}{9}               & \multicolumn{1}{r|}{5}               & \multicolumn{1}{r|}{0}               & \multicolumn{1}{r|}{15}              & \multicolumn{1}{r|}{14}              & \multicolumn{1}{r|}{2}               & \multicolumn{1}{r|}{3}               & \multicolumn{1}{r|}{12}              \\ \hline
\multicolumn{17}{c}{\textbf{}}                                                                                                                                                                                                                                                                                                                                                                                                                                                                                                                                                                                                                                                        \\ \hline
\multicolumn{1}{|r|}{\textbf{S8}}     & \multicolumn{1}{r|}{\textbf{x0000x}} & \multicolumn{1}{r|}{\textbf{x0001x}} & \multicolumn{1}{r|}{\textbf{x0010x}} & \multicolumn{1}{r|}{\textbf{x0011x}} & \multicolumn{1}{r|}{\textbf{x0100x}} & \multicolumn{1}{r|}{\textbf{x0101x}} & \multicolumn{1}{r|}{\textbf{x0110x}} & \multicolumn{1}{r|}{\textbf{x0111x}} & \multicolumn{1}{r|}{\textbf{x1000x}} & \multicolumn{1}{r|}{\textbf{x1001x}} & \multicolumn{1}{r|}{\textbf{x1010x}} & \multicolumn{1}{r|}{\textbf{x1011x}} & \multicolumn{1}{r|}{\textbf{x1100x}} & \multicolumn{1}{r|}{\textbf{x1101x}} & \multicolumn{1}{r|}{\textbf{x1110x}} & \multicolumn{1}{r|}{\textbf{x1111x}} \\ \hline
\multicolumn{1}{|r|}{\textbf{0yyyy0}} & \multicolumn{1}{r|}{13}              & \multicolumn{1}{r|}{2}               & \multicolumn{1}{r|}{8}               & \multicolumn{1}{r|}{4}               & \multicolumn{1}{r|}{6}               & \multicolumn{1}{r|}{15}              & \multicolumn{1}{r|}{11}              & \multicolumn{1}{r|}{1}               & \multicolumn{1}{r|}{10}              & \multicolumn{1}{r|}{9}               & \multicolumn{1}{r|}{3}               & \multicolumn{1}{r|}{14}              & \multicolumn{1}{r|}{5}               & \multicolumn{1}{r|}{0}               & \multicolumn{1}{r|}{12}              & \multicolumn{1}{r|}{7}               \\ \hline
\multicolumn{1}{|r|}{\textbf{0yyyy1}} & \multicolumn{1}{r|}{1}               & \multicolumn{1}{r|}{15}              & \multicolumn{1}{r|}{13}              & \multicolumn{1}{r|}{8}               & \multicolumn{1}{r|}{10}              & \multicolumn{1}{r|}{3}               & \multicolumn{1}{r|}{7}               & \multicolumn{1}{r|}{4}               & \multicolumn{1}{r|}{12}              & \multicolumn{1}{r|}{5}               & \multicolumn{1}{r|}{6}               & \multicolumn{1}{r|}{11}              & \multicolumn{1}{r|}{0}               & \multicolumn{1}{r|}{14}              & \multicolumn{1}{r|}{9}               & \multicolumn{1}{r|}{2}               \\ \hline
\multicolumn{1}{|r|}{\textbf{1yyyy0}} & \multicolumn{1}{r|}{7}               & \multicolumn{1}{r|}{11}              & \multicolumn{1}{r|}{4}               & \multicolumn{1}{r|}{1}               & \multicolumn{1}{r|}{9}               & \multicolumn{1}{r|}{12}              & \multicolumn{1}{r|}{14}              & \multicolumn{1}{r|}{2}               & \multicolumn{1}{r|}{0}               & \multicolumn{1}{r|}{6}               & \multicolumn{1}{r|}{10}              & \multicolumn{1}{r|}{13}              & \multicolumn{1}{r|}{15}              & \multicolumn{1}{r|}{3}               & \multicolumn{1}{r|}{5}               & \multicolumn{1}{r|}{8}               \\ \hline
\multicolumn{1}{|r|}{\textbf{1yyyy1}} & \multicolumn{1}{r|}{2}               & \multicolumn{1}{r|}{1}               & \multicolumn{1}{r|}{14}              & \multicolumn{1}{r|}{7}               & \multicolumn{1}{r|}{4}               & \multicolumn{1}{r|}{10}              & \multicolumn{1}{r|}{8}               & \multicolumn{1}{r|}{13}              & \multicolumn{1}{r|}{15}              & \multicolumn{1}{r|}{12}              & \multicolumn{1}{r|}{9}               & \multicolumn{1}{r|}{0}               & \multicolumn{1}{r|}{3}               & \multicolumn{1}{r|}{5}               & \multicolumn{1}{r|}{6}               & \multicolumn{1}{r|}{11}              \\ \hline
\end{tabular}}
\end{table}

\subsection{Key generation permutation tables}

\begin{table}[H]
\caption{Key Permutation ($PC1$). Note that only 56 of the 64 bits are used (the remaining are parity bits).}
\label{tbl:des-PC1}
\begin{tabular}{|lllllll||lllllll|}
\hline
\multicolumn{7}{|c||}{Left}                                                                                                                                      & \multicolumn{7}{c|}{Right}                                                                                                                                     \\ \hline
\multicolumn{1}{|l|}{57} & \multicolumn{1}{l|}{49} & \multicolumn{1}{l|}{41} & \multicolumn{1}{l|}{33} & \multicolumn{1}{l|}{25} & \multicolumn{1}{l|}{17} & 9  & \multicolumn{1}{l|}{63} & \multicolumn{1}{l|}{55} & \multicolumn{1}{l|}{47} & \multicolumn{1}{l|}{39} & \multicolumn{1}{l|}{31} & \multicolumn{1}{l|}{23} & 15 \\ \hline
\multicolumn{1}{|l|}{1}  & \multicolumn{1}{l|}{58} & \multicolumn{1}{l|}{50} & \multicolumn{1}{l|}{42} & \multicolumn{1}{l|}{34} & \multicolumn{1}{l|}{26} & 18 & \multicolumn{1}{l|}{7}  & \multicolumn{1}{l|}{62} & \multicolumn{1}{l|}{54} & \multicolumn{1}{l|}{46} & \multicolumn{1}{l|}{38} & \multicolumn{1}{l|}{30} & 22 \\ \hline
\multicolumn{1}{|l|}{10} & \multicolumn{1}{l|}{2}  & \multicolumn{1}{l|}{59} & \multicolumn{1}{l|}{51} & \multicolumn{1}{l|}{43} & \multicolumn{1}{l|}{35} & 27 & \multicolumn{1}{l|}{14} & \multicolumn{1}{l|}{6}  & \multicolumn{1}{l|}{61} & \multicolumn{1}{l|}{53} & \multicolumn{1}{l|}{45} & \multicolumn{1}{l|}{37} & 29 \\ \hline
\multicolumn{1}{|l|}{19} & \multicolumn{1}{l|}{11} & \multicolumn{1}{l|}{3}  & \multicolumn{1}{l|}{60} & \multicolumn{1}{l|}{52} & \multicolumn{1}{l|}{44} & 36 & \multicolumn{1}{l|}{21} & \multicolumn{1}{l|}{13} & \multicolumn{1}{l|}{5}  & \multicolumn{1}{l|}{28} & \multicolumn{1}{l|}{20} & \multicolumn{1}{l|}{12} & 4  \\ \hline
\end{tabular}
\end{table}

\begin{table}[H]
\caption{Key Permutation ($PC2$). Note that the input is the two 28-bit key halves concatenated.}
\label{tbl:des-PC2}
\begin{tabular}{|r|r|r|r|r|r|}
\hline
14 & 17 & 11 & 24 & 1  & 5  \\ \hline
3  & 28 & 15 & 6  & 21 & 10 \\ \hline
23 & 19 & 12 & 4  & 26 & 8  \\ \hline
16 & 7  & 27 & 20 & 13 & 2  \\ \hline
41 & 52 & 31 & 37 & 47 & 55 \\ \hline
30 & 40 & 51 & 45 & 33 & 48 \\ \hline
44 & 49 & 39 & 56 & 34 & 53 \\ \hline
46 & 42 & 50 & 36 & 29 & 32 \\ \hline
\end{tabular}
\end{table}

\begin{table}[H]
\caption{Key bit rotations.}
\label{tbl:des-SHIFTS}
\begin{tabular}{|c|l|l|l|l|l|l|l|l|l|l|l|l|l|l|l|l|}
\hline
\textbf{Round Number}             & 1 & 2 & 3 & 4 & 5 & 6 & 7 & 8 & 9 & 10 & 11 & 12 & 13 & 14 & 15 & 16 \\ \hline
\textbf{Number of Left Rotations} & 1 & 1 & 2 & 2 & 2 & 2 & 2 & 2 & 1 & 2  & 2  & 2  & 2  & 2  & 2  & 1  \\ \hline
\end{tabular}
\end{table}

\section{Formula for the SAT Attack Worked Example}
The formula below should return UNSAT when used with a SAT solver. Removing the \texttt{(y + yx)} part should make the formula SAT and reveal the key. 

\begin{lstlisting}[breaklines=true,basicstyle=\tiny]
((~(a&b) <=> u0) & (~(b&c) <=> u1) & ((c&a) <=> u2) & (~(k1a+u0) <=> u5) & ((k2a + u1) <=> u4) & (~(k3a+u2) <=> u3) & (~(u5&u4) <=> u6) &
(y <=> (u6 | u3)))
&
((~(a&b) <=> w0) & (~(b&c) <=> w1) & ((c&a) <=> w2) & (~(k1b+w0) <=> w5) & ((k2b + w1) <=> w4) & (~(k3b+w2) <=> w3) & (~(u5&w4) <=> w6) &
(yx <=> (w6 | w3)))
&
(y + yx) &

(
((~(aa&bb) <=> uu0) & (~(bb&cc) <=> uu1) & ((cc&aa) <=> uu2) & (~(k1a+uu0) <=> uu5) & ((k2a + uu1) <=> uu4) & (~(k3a+uu2) <=> uu3) & (~(uu5&uu4) <=> uu6) &
(yy <=> (uu6 | uu3)))
&
((~(aa&bb) <=> ww0) & (~(bb&cc) <=> ww1) & ((cc&aa) <=> ww2) & (~(k1b+ww0) <=> ww5) & ((k2b + ww1) <=> ww4) & (~(k3b+ww2) <=> ww3) & (~(uu5&ww4) <=> ww6) &
(yy <=> (ww6 | ww3)))
&
(aa &
~bb &
cc &
yy )
)

&

(
((~(aaa&bbb) <=> uuu0) & (~(bbb&ccc) <=> uuu1) & ((ccc&aaa) <=> uuu2) & (~(k1a+uuu0) <=> uuu5) & ((k2a + uuu1) <=> uuu4) & (~(k3a+uuu2) <=> uuu3) & (~(uuu5&uuu4) <=> uuu6) &
(yyy <=> (uuu6 | uuu3)))
&
((~(aaa&bbb) <=> www0) & (~(bbb&ccc) <=> www1) & ((ccc&aaa) <=> www2) & (~(k1b+www0) <=> www5) & ((k2b + www1) <=> www4) & (~(k3b+www2) <=> www3) & (~(uuu5&www4) <=> www6) &
(yyy <=> (www6 | www3)))
&
(~aaa &
bbb &
ccc &
yyy )
)

&

(
((~(aaaa&bbbb) <=> uuuu0) & (~(bbbb&cccc) <=> uuuu1) & ((cccc&aaaa) <=> uuuu2) & (~(k1a+uuuu0) <=> uuuu5) & ((k2a + uuuu1) <=> uuuu4) & (~(k3a+uuuu2) <=> uuuu3) & (~(uuuu5&uuuu4) <=> uuuu6) &
(yyyy <=> (uuuu6 | uuuu3)))
&
((~(aaaa&bbbb) <=> wwww0) & (~(bbbb&cccc) <=> wwww1) & ((cccc&aaaa) <=> wwww2) & (~(k1b+wwww0) <=> wwww5) & ((k2b + wwww1) <=> wwww4) & (~(k3b+wwww2) <=> wwww3) & (~(uuuu5&wwww4) <=> wwww6) &
(yyyy <=> (wwww6 | wwww3)))
&
(~aaaa &
bbbb &
~cccc &
~yyyy )
)
\end{lstlisting}

\end{document}